\documentclass[aps,prd,showpacs,superscriptaddress,groupedaddress]{revtex4-2}
\usepackage[T1]{fontenc}
\usepackage{lmodern} 
\usepackage{graphicx}  % needed for figures
\usepackage{dcolumn}   % needed for some 
\usepackage{bm}        % for math
\usepackage{amssymb}   % for math
\usepackage{amsmath}
\usepackage{mathtools}
\usepackage{bbold}
\usepackage{array}
\usepackage{makecell}
\usepackage{braket}
\usepackage{longtable}
\usepackage{supertabular,booktabs}
\usepackage[mathscr]{euscript}
\usepackage{xcolor}
\usepackage{soul}
\usepackage{titlesec} %for hyperref to detect names of section etc
\usepackage[colorlinks,citecolor=blue,urlcolor=blue,hypertexnames=true]{hyperref}
\usepackage{subfigure}

\newcommand{\be}{\begin{equation}}
\newcommand{\ee}{\end{equation}}
\newcommand{\bea}{\begin{eqnarray}}
\newcommand{\eea}{\end{eqnarray}}
\newcommand{\bml}{\begin{subequations}}
\newcommand{\eml}{\end{subequations}}
\newcommand{\bfig}{\begin{figure}}
\newcommand{\efig}{\end{figure}}

\newcommand{\der}{\partial}

\newcommand{\bmat}{\begin{pmatrix}}
\newcommand{\emat}{\end{pmatrix}}
\usepackage{graphicx,slashed,booktabs,xcolor,multirow,float,amsfonts,bbold,mathtools,sidecap,tikz,bm,enumitem}

\definecolor{lime}{HTML}{A6CE39}
\DeclareRobustCommand{\orcidicon}{\hspace{-2.1mm}
\begin{tikzpicture}
\draw[lime,fill=lime] (0,0.0) circle [radius=0.13] node[white] {{\fontfamily{qag}\selectfont \tiny \,ID}}; \draw[white, fill=white] (-0.0525,0.095) circle [radius=0.007]; 
\end{tikzpicture} \hspace{-3.7mm} }
\foreach \x in {A, ..., Z}{\expandafter\xdef\csname orcid\x\endcsname{\noexpand\href{https://orcid.org/\csname orcidauthor\x\endcsname} {\noexpand\orcidicon}}}

\begin{document}
% The following information is for internal review, please remove them for submission
\widetext

% the following line is for submission, including submission to the arXiv!!
%\hspace{5.2in} \mbox{Fermilab-Pub-04/xxx-E}

\title{\textcolor{blue}{\fontsize{21}{21}\textbf{Schwinger-Keldysh path integral formalism
for a Quenched Quantum Inverted Oscillator}}}
%\input author_list.tex       % D0 authors (remove the first 3 lines
                             % of this file prior to submission, they
                             % contain a time stamp for the authorlist)
                             % (includes institutions and visitors)
%\date{\today}

%\date{\today}
\author{{\large  Sayantan Choudhury\orcidA{}${}^{1}$}}
\email{Corresponding Author: 
sayantan_ccsp@sgtuniversity.org,  sayanphysicsisi@gmail.com}

\author{ \large Suman Dey${}^{2}$}
\email{dey.suman.vbu@gmail.com}
\author{ \large Rakshit Mandish Gharat${}^{3}$}
\email{rakshitmandishgharat.196ph018@nitk.edu.in}
\author{\large Saptarshi Mandal${}^{4}$}
\email{saptarshijhikra@gmail.com}
%\author{\large Silpadas N${}^{5}$}
%\email{\textcolor{black}{silpadasn@gmail.com }}
%\author{\large Sudhakar Panda ${}^{6,7}$}
%\email{\textcolor{black}{panda@niser.ac.in}}
\author{\large Nilesh Pandey${}^{1}$}
\email{nilesh911999@gmail.com}

\affiliation{ ${}^{1}$Centre For Cosmology and Science Popularization (CCSP),
SGT University, Gurugram, Delhi- NCR, Haryana- 122505, India,}
\affiliation{${}^{2}$ Department of Physics, Visva-Bharati University, Santiniketan, Birbhum-731235, India,}
\affiliation{${}^{3}$Department of Physics, National Institute of Technology Karnataka, Surathkal, Karnataka-575025, India,}
\affiliation{${}^{4}$ Department of Physics, Indian Institute of Technology Kharagpur, Kharagpur-721302, India,}
%\affiliation{${}^{5}$ Department of Physics, Pondicherry University, India,}
%\affiliation{${}^{6}$National Institute of Science Education and Research, Jatni, Bhubaneswar, Odisha - 752050, India,}
%\affiliation{${}^{7}$Homi Bhabha National Institute, Training School Complex, Anushakti Nagar, Mumbai -400085, India,}
%\affiliation{${}^{5}$ Department of Applied Physics, Delhi Technological University, Delhi-110042, India.}

\begin{abstract}
In this work, we study the time-dependent behaviour of quantum correlations of a system of an inverted oscillator governed by out-of-equilibrium dynamics using the well-known {\it Schwinger-Keldysh formalism} in presence of quantum mechanical quench.  Considering a generalized structure of a time-dependent Hamiltonian for an inverted oscillator system,  we use the {\it invariant operator method} to obtain its eigenstate and continuous energy eigenvalues.  Using the expression for the eigenstate,  we further derive the most general expression for the generating function as well as the out-of-time-ordered correlators (OTOC) for the given system using this formalism. Further,  considering the time-dependent coupling and frequency of the quantum inverted oscillator characterized by quench parameters, we comment on the dynamical behaviour,  specifically the early,  intermediate and late time-dependent features of the OTOC for the quenched quantum inverted oscillator.  Next,  we study a specific case,  
where the system of inverted oscillator exhibits chaotic behaviour by computing the {\it quantum Lyapunov exponent} from the time-dependent behaviour of OTOC  in presence of the given quench profile.

\end{abstract}

%\pacs{}

\maketitle

\section{\textcolor{blue}{\textbf{ \large  Introduction}}}
\label{sec:introduction}
%%sk path non-equilibrium, OTOC chaos, inverted oscillator, quench, motivation
In theoretical high energy physics, the underlying concept of the Feynman path integral is applicable to quantum systems for which the vacuum state in far future is exactly the same as in far past. Clearly, the Feynman path integral is not applicable for quantum systems with a dynamical background.
The Schwinger-Keldysh \cite{1961JMP.....2..407S, keldysh1965diagram,SK1,SK2,SK3} formalism appears within the framework of quantum mechanics, where the physical system is placed on a dynamical background, i.e., {quantum dynamics of the system is far from equilibrium.} Schwinger-Keldysh formalism seems to be the most formidable choice to evaluate generating functionals and correlation functions in out-of-equilibrium quantum field theory and statistical mechanics \cite{BenTov:2021jsf}. It unravels the prodigious toolbox of equilibrium quantum field theory to out-of-equilibrium problems. There are large numbers of applications of out-of-equilibrium dynamics in various fields, for instance, condensed matter physics\cite{2019arXiv190507403B,Sieberer:2015hba,Georgii+2011,1957NCim....6..371T,9}, cosmology \cite{Choudhury:2018rjl,Akhtar:2019qdn,Chen:2017ryl, Boyanovsky:1993xf} (in the context of the primordial universe), black hole physics\cite{Glorioso:2018mmw,Haehl:2015foa}, high energy physics - theory and phenomenology \cite{Herzog:2002pc,Giecold:2009tt,Choudhury:2017bou,Choudhury:2017qyl,2022PhRvD.106b5002C}, and more.

The time-dependent states for dynamical quantum systems can be computed by solving \textit{Time-Dependent Schrodinger Equation} (TDSE). One of the ways to solve TDSE is by constructing the \textit{Lewis-Resenfield invariant operator} and this is often termed as invariant operator representation of the wavefunction \cite{doi:10.1063/1.1664991}. Some works following this approach to compute the time-dependent eigenstates are reported in refs. \cite{intro:invar1,intro:invar2,invarSC,intro:invar3,intro:invar4}. 

Highly complex behavior of quantum systems can often be studied using a simple model of Standard Harmonic Oscillator (SHO). Sister to SHO is the Inverted Harmonic Oscillator (IHO) that has remarkable properties which are applicable throughout many disciplines of physics \cite{Choudhury:2011jt,Choudhury:2012yh,Choudhury:2011sq,Choudhury:2013zna,Choudhury:2015hvr,Bhattacharyya:2020kgu,Choudhury:2017cos,Akhtar:2019qdn,Choudhury:2016pfr,Choudhury:2016cso,Ali:2019zcj,Bhattacharyya:2020rpy,Choudhury:2017bou,Einhorn:2003xb,Grain:2019vnq,Grishchuk:1992tw,Bhargava:2020fhl,Choudhury:2020hil,Adhikari:2021pvv,Choudhury:2021brg,Martin:2021qkg,Choudhury:2022btc}. The quantum treatment of IHO is exactly solvable similar to that of SHO. Unlike SHO, IHO can be used to model complicated types of out-of-equilibrium systems.

{A key idea for quantifying the propagation of detailed quantum correlation at out-of-equillbrium, information theoretic scrambling via exponential growth, and quantum chaos can be easily studied via out-of-time-ordered correlators or simply "OTOC" \cite{24}.} OTOC has been widely used as a tool to probe quantum chaos in quantum systems \cite{OTOC1,OTOC2,OTOC3,OTOC4,Choudhury:2020yaa,OTOC5,OTOC6,OTOC7}. In a more general sense OTOC is a mathematical tool using which one can easily quantify quantum mechanical correlation at out-of-equilibrium. Motivated by the work of \cite{Shenker:2013pqa}, Kitaev \cite{20} emphasized arbitrarily ordered 4-point correlation functions, as "out-of-time-ordered correlators", in constructing analogies between effective field theories, making the connection between superconductors \cite{1969JETP...28.1200L} and black holes \cite{dray1985gravitational, tHooft:1990fkf}.

Most of the recent works in many-body physics focus on studying dynamics of quantum systems where some time-dependent parameter is varied suddenly or very slowly. This is commonly refereed to as a Quantum Quench. The quench protocol is said to drive any out of equilibrium system and in turn can {trigger the thermalisation process of} these systems, \cite{intro:ooe1,intro:ooe2,intro:ooe3}.  The effects of quenches in quantum systems have even been explored experimentally for cold atoms in \cite{e1,e2,e3,e4,e5,e6,e7,e8,e9,e10}. A few of the works focusing on OTOCs in the context of quenched systems are \cite{OTOCq1,OTOCq2,OTOCq3,OTOCq4,OTOCq5}.

Motivated by all these ideas, in this work, we compute the OTOC for inverted oscillators having a quenched coupling and frequency, using Schwinger-Keldysh path integral formalism. In section \ref{genHam} the continuous energy eigenvalues and normalized wavefunction for a generic Hamiltonian of a time-dependent inverted oscillator are computed  \cite{doi:10.1139/cjp-2014-0553} using the Lewis–Riesenfeld invariant method. In section (\ref{claction}), we construct the action for the Lagrangian of generic time-dependent inverted oscillator. Using this action we derive the generating function for the inverted oscillator in section \ref{genfuncsk}. We modify this generating function using Shwinger-Keldysh path integral formalism in section \ref{SKpath}, following the prescription of \cite{BenTov:2021jsf}. Using this generating function and the Green's functions computed in Appendix \ref{greenf} we finally compute the OTOC for inverted oscillator in terms of Green's functions in section \ref{otocsec}. In the section (\ref{sec:numerical}), we have numerically studied dynamical behaviour of the OTOC by quenching coupling and frequency of the inverted oscillator. We have also computed the Lyapunov exponent and commented on the chaotic behavior of the inverted oscillator in this section. Section (\ref{discuss}) serves as the final section of our study before providing pertinent future possibilities and directions.

\section{\textcolor{blue}{\textbf{\large Formulation of Time-dependent Generalized Hamiltonian Dynamics}}}\label{genHam}
%%First Action. Generating. SK path, Influence phase, 
\noindent In this section, we explicitly discuss the crucial role of a system {described by} a time-dependent {quantum mechanical} inverted oscillator in a very generic way to deal with path integrals and {different types of quantum correlation functions (anti time ordered, time ordered, out-of-time-ordered)}. A generalised Hamiltonian for a time-dependent inverted oscillator can be written as:
\begin{equation}\label{IO}
    \hat{H}(t) = \frac{1}{m(t)} \hat{p}^{2} -\frac{1}{2} m(t) \Omega^{2}(t)\hat{q}^{2} + \frac{1}{2} f(t)\left(\hat{p}\hat{q}+\hat{q}\hat{p}\right).
\end{equation}
\noindent Here $\hat{q}(t)$ denotes the time-dependent generalized coordinate, $m(t)$ is the time-dependent mass, $\Omega (t)$ is the time-dependent frequency, and $f(t)$ is a time-dependent coupling parameter. {In general these time dependent parameters can be anything}. It is significant to note that, in contrast to the {SHO}, the second term, which represents the potential energy of the inverted harmonic oscillator, has a negative sign. The main reason for the negative sign in the potential energy is that for inverted oscillator, the harmonic oscillator frequency, $\omega (t)$, is replaced by $i\omega (t)$. We can write the Euler-Lagrange equation for the coordinate $\hat{q}(t)$ of the inverted oscillator as:
\begin{equation}\label{EL}
    \left(\mathcal{D}_{t}-\Omega_{eff}^{2}\right)\hat{q}(t)=0,
\end{equation}
\noindent where the differential operator $\mathcal{D}_{t}$ is defined as:
\begin{equation}\label{de1}
    \mathcal{D}_{t}=\left(\frac{d^{2}}{dt^{2}}+\frac{d~ \text{ln}~m(t)}{dt}\frac{d}{dt}\right).
\end{equation}
The squared effective time-dependent frequency in Eq.\eqref{EL} is defined by:
\begin{equation}
\Omega_{eff}^{2}(t) = \Omega^{2}(t)+f^{2}(t)+\frac{\dot{m}(t)}{m(t)} f(t) + \dot{f}(t)\label{ome12}.
\end{equation}
{where the dot corresponds to the differentiation with respect to time "t"}.
\noindent Then, the Euler-Lagrange equation in \eqref{de1} gives the equation of motion in the simplified form as:
\begin{equation}
    \ddot{\hat{q}}+\frac{\dot{m}}{m}\dot{\hat{q}}-\Omega_{eff}^{2} (t) \hat{q}=0.
\end{equation}
\noindent Next, the TDSE for {the above mentioned time dependent system} can then be written as:
\begin{equation}\label{ham}
    \hat{H}(t)\Psi(q,t)=i \frac{\partial \Psi(q,t)}{\partial t}.
\end{equation}
{Now}, to derive the eigenstates and energy eigenvalues of the above TDSE, we will use Lewis-Resenfield invariant operator method, in the next section, {which will be an extremely useful tool to deal with the rest of the problem.} 

\subsection{\textcolor{blue}{\textbf{\normalsize Lewis–Riesenfeld Invariant Method}}}
In this subsection we apply the technique of Lewis-Riesenfield invariant operator method to calculate the time dependent eigenstates and energy eigenvalues for the generalised 
Hamiltonian of inverted oscillator {as} given in Eq.\eqref{IO}.

\noindent The Lewis-Riesenfield invariant method is advantageous for obtaining the complete solution of an inverted harmonic oscillator with a time-dependent frequency having $\mathcal{P}\mathcal{T}$ symmetry. In this subsection, we will construct a dynamical invariant operator to transit the eigenstates of the Hamiltonian from prescribed initial to final configurations, in arbitrary time. Let us assume an Invariant operator $\hat{I}(t)$ which is a hermitian operator and explicitly satisfies the relation:
\begin{equation}\label{5}
    \frac{d\hat{I}(t)}{dt}=\frac{1}{i}[\hat{I}(t), \hat{H}(t)]+\frac{\partial \hat{I}(t)}{\partial t},
\end{equation}
\noindent so that its expectation values remain constant in time. Here $\hat{H}(t)$ is the time-dependent Hamiltonian of our system defined earlier in Eq.\eqref{IO}. If the exact form of invariant operator, $\hat{I}(t)$, does not contain any time derivative operators, it allows one to write the solutions of the {TDSE} as:
\begin{equation}\label{LReigen}
    \psi_{n}(q,t)=e^{i \mu_{n}(t)} \varphi_{n}(q,t).
\end{equation}
\noindent Here, $\varphi_{n}(q,t)$ is an eigenfunction of $\hat{I}(t)$ with  eigenvalue $n$ and $\mu_{n}(t)$ is the {time-dependent phase function}. Note that for the sake of simplicity, we are writing $\hat{q}$ as $q$ and the same for other operators. {This means we further remove all hats}.
\noindent Next, we consider that time-dependent linear invariant operator for the system in the form:
\begin{equation}\label{8}
    I(t)=\alpha(t)q + K(t)p+\gamma(t),
\end{equation}
\noindent where $\alpha(t)$, $K(t)$, and $\gamma(t)$ are the time-dependent real functions. Using Eq.\eqref{5} we get:
\begin{equation}\label{9}
 \begin{aligned}
    \dot{\alpha}(t)+\alpha(t)f(t) = -K(t)m(t)\Omega^{2}(t)\\
    \dot{K}(t)-K(t)f(t)=-\frac{\alpha(t)}{m(t)}~~~~~~~~~~\\
    \dot{\gamma}(t)=0.~~~~~~~~~~~~~~~~~~~~
 \end{aligned}
\end{equation}
\noindent From Eq. (\ref{9}) one can find the following second order differential equation:
\begin{equation}
    \ddot{K}+\frac{\dot{m}}{m}\dot{K}-\Omega_{eff}^{2}(t)K =0,
\end{equation}
\noindent where the squared effective time-dependent frequency is defined in Eq.\eqref{ome12} before. Therefore, we can write the Eq.\eqref{8} as:
\begin{equation}\label{Inv}
    I(t) = K(t) p-m(t)\left[\dot{K(t)}-K(t)f(t)\right]q.
\end{equation}
\noindent Now we intend to find the eigenstate $\varphi_{n}(q,t)$ of $I(t)$ {using the following eigenvalue equation:} 
\begin{equation}\label{Inv2}
    I\varphi_{n}(q,t)=n\varphi_{n}(q,t)
\end{equation}
{Here, the eigenstates satisfy the orthonormalization condition,which is,} $\langle \varphi_{n}|\varphi_{n'}\rangle=\delta(n-n')$. Subsituting the expression of $I(t)$ from Eq. \eqref{Inv} in the above Eq.\eqref{Inv2}, one can find the continuous eigenstates of the inverted oscillator:  
\begin{equation}
\varphi_{n}=N \exp{\Bigg\{\frac{im(t)}{2K(t)}\biggr[[\dot{K}(t)-f(t)K(t)]q^2+\frac{2n}{m(t)}q\biggr]\Bigg\}}.
\end{equation}
It is very straightforward to calculate the normalizing constant, {N, which is given by the following expression:} 
\begin{equation}
   N = \sqrt{\frac{1}{2\pi K}}.
 \end{equation}
One can find the phase factor by inserting the eigenfunction of invariant operator in TDSE and solving the {following} equation:
\begin{equation}\dot{\mu}_{n}\varphi_{n}=\left(i \frac{\partial}{\partial t}-H\right)\varphi_{n}.%\hspace*{1.0cm}
\end{equation}

Using Eq.\eqref{LReigen}, we can write the wavefunction in the normalized form:
\be
\psi_{n}=\sqrt{\frac{1}{2\pi K}}~e^{i\mu_{n}(t)}\exp{\Bigg\{\frac{im(t)}{2K(t)}\biggr[[\dot{K}(t)-f(t)K(t)]q^2+\frac{2n}{m(t)}q\biggr]\Bigg\}}.
\ee

Now, to prove that the wave functions are always finite, we have to find the integration over all possible continuous eigenvalues, {which is given by:}
\be\label{psiuf}
\Psi(q,t)=\int_{-\infty}^{\infty}g(n)~\psi_n(q,t)~dn.
\ee
\noindent Here $g(n)$ is the weight function. It helps one to find which state the system is in. Let's consider the weight function as {a gaussian function, which is of the following exact form:}
\be
g(n)=\frac{\sqrt{a}}{(2\pi)^{\frac{1}{4}}}\exp{\biggr(-\frac{a^{2}n^{2}}{4}\biggr)}.
\ee
Here `$a$' is a real positive constant. Substituting the above form of weight function in Eq.\eqref {psiuf} and integrating over all possible continuous eigenvalues, one can write the wavefunction for inverted oscillator as:
\be\label{psim}
\Psi(q,t)=\left(\frac{2A(t)}{\pi}\right)^{1/4}\exp{\biggl\{iB_{2}(t)+[iB_{1}(t)-A(t)]q^{2}\biggr\}}.
\ee
Here,{we define the time dependent functions as:}
\begin{flalign}
    &\nonumber A(t)=\biggr[K^{2}a^{2}\biggr(1-\frac{4y^{2}}{a^{4}}\biggr)\biggr]^{-1},\\
    &\nonumber B_{1}(t)=\frac{m(\dot{K}-fK)}{2K}+\frac{2y}{K^{2}a^{4}\biggr(1+\frac{4y^{2}}{a^{4}}\biggr)},\\
    &\nonumber B_{2}(t)=\frac{1}{2}\tan^{-1}{\frac{2y}{a^{2}}},\\
    & y(t)=\int_{0}^{t}\frac{d\tau}{m(\tau)K^{2}(\tau)}.
\end{flalign}
\noindent Now having obtained the normalization factor, we can write the {expression for the} energy of our desired quantum system as:
\be\label{Energy}
E=\sqrt{\frac{2A(t)}{\pi}}\int_{-\infty}^{\infty}\Psi^{*}\hat{H}\Psi~dq.
\ee
Inserting the Eq.\eqref{psim}, $\hat{H}$ and conjugate of Eq.\eqref{psim} into the Eq.\eqref{Energy} one can show that the time-dependent energy for inverted oscillator is {given by the following expression:},
\be
E=\frac{1}{2m(t)}\biggr(\frac{B_{1}^{2}(t)}{A(t)}+A(t)\biggr)+\frac{1}{8}\frac{m(t)\Omega^{2}(t)}{A(t)}+\frac{f(t)B_{1}(t)}{2A(t)}.
\ee
\noindent It is interesting to observe that the energy eigenvalues are independent of $n$ but energy is a continuous function of time. Furthermore, there is no zero-point energy associated with this system of inverted oscillator.
\section{\textcolor{blue}{\textbf{\large Evaluation of Action}}}\label{claction}

In this section we start with a Lagrangian of the inverted oscillator and derive the {representative} action for the generalised inverted oscillator. We express the action in terms of Green's functions which are derived in Appendix \ref{greenf}. The action thus obtained will be useful for deriving the generating function of inverted oscillator. {Here, it is important to note that the generating function in the present context physically represents the partition function in Euclidean signature.}

As the Green's function for inverted oscillator, given in Eq.\eqref{gfhyp} is hyperbolic in nature, one can use the hyperbolic identities. Inferring from \cite{BenTov:2021jsf} one can then write the classical field solution for the inverted oscillator as, 
\be\label{classfi}
\bar{q}(t)=\frac{1}{G(T)}\biggr\{q_f G(t-t_0)+q_0~G(t_f-t)+\int_{t_0}^{t_f}dt'\left[G(T)\Theta(t-t')G(t-t')-G(t-t_0)G(t_f-t')\right]J(t')\biggr\}.
\ee
The time-derivative of this field can be given by,
\be
\dot{\bar{q}}=\frac{1}{G(T)} \biggr[q_{f}\dot{G}(t-t_0)-q_0 \dot{G}(t-t_f)+\int_{t_0}^{t_f}dt' [G(T)\Theta(t-t')\dot{G}(t-t')-\dot{G}(t-t_0)G(t_f-t')]J(t')\biggr].
\ee
Here $T\equiv t_f - t_0$, such that $t_0$ and $t_f$ denote the initial time and final time respectively. Also, $\bar{q}(t_0)=q_0$ and $\bar{q}(t_f)=q_f$ are the intial and final field configurations respectively. Applying the boundary condition $\dot{G}(0)=1$ one can find,
\be\label{qdi}
\dot{\bar{q}}(t_0)=\frac{1}{G(T)}\biggr[q_f-q_0\dot{G}(T)-\int_{t_0}^{t_f}dt'~G(t_f-t')J(t')\biggr].
\ee
Furthermore, using the hyperbolic identity,
\be
\dot{G}(T)G(t_f-t)-\dot{G}(T)G(t_f-t)=G(t-t_0),
\ee
we obtain:
\be\label{qdf}
\dot{\bar{q}}(t_f)=\frac{1}{G(T)}\biggr[q_f\dot{G}(T)-q_0+\int_{t_0}^{t_f}dt'~G(t'-t_0)J(t')\biggr]
\ee

In general, the action for any field is given by: 
\begin{equation}\label{actionfe}
    S = \int_{t_0}^{t_f} [\mathcal{L}+J(t)q(t)] dt.
\end{equation}
Here the term $J(t)$ is an auxiliary {time dependent} field and $\mathcal{L}$ is the Lagrangian of the system.
For an inverted oscillator the { the representative Lagrangian can be written as:}
\begin{equation}\label{lgrng}
    \mathcal{L}=\frac{1}{2} m(t)\dot{q}^2+m(t)\omega(t)^2q^2-f(t)m(t)\dot{q}q
\end{equation}
Substituting, Eq.\eqref{lgrng} in Euler Lagrange equation the equation of motion for the inverted oscillator becomes,
\be\label{EOM}
m(t)\ddot{q}+\dot{m(t)}\dot{q}-\Bigg[m(t)\omega(t)^2-f(t)\dot{m(t)}-m(t)\dot{f(t)}\Bigg]q=0
\ee
Substituting Eq.\eqref{lgrng} in \eqref{actionfe} the action for the inverted oscillator { can be expressed as:}
\be 
S = \int_{t_0}^{t_f} \biggr[\bigg(\frac{1}{2} m(t)\dot{q}^2+\frac{1}{2}m(t)\omega^2(t)q^2-f(t)m(t)\dot{q}q\bigg)+J(t)q(t)\biggr] dt.
\ee
 
Using the equation of motion i.e. Eq.\eqref{EOM} and the definition of classical field solution {as stated} in Eq.\eqref{classfi}, we find the classical limit for the action of inverted oscillator as,
\be\label{action limit}
S_{cl} (q_0,q_f|J) \equiv S(\bar{q}) = \frac{1}{2}  m(t)\bar{q}\dot{\bar{q}}\biggr|_{t_0}^{t_f}-\frac{1}{2} f(t)m(t)\bar{q}^{2}\biggr|_{t_0}^{t_f}+\frac{1}{2} \int_{t_0}^{t_f} dt J(t)\bar{q}(t).
\ee
The final expression for the action can be obtained using Eq.\eqref{qdi} and Eq.\eqref{qdf}, as given below:
\be  
\begin{split}\label{action}S(q_0&,q_f|J)=\frac{1}{2} \biggr[m(t_f)\frac{1}{G(T)}\biggr\{q_f^2\dot{G}(T)-q_0q_f+\int_{t_0}^{t_f}dt'q_fG(t'-t_0)J(t')\biggr\}-m(t_0)\frac{1}{G(T)}\biggr\{q_fq_0-{q_0^2}\dot{G}(T)+\\&\int_{t_0}^{t_f}dt'q_0G(t_f-t')J(t')\biggr\}\biggr]-\frac{1}{2}\biggr[f(t_f)m(t_f)q_f^2-f(t_0)m(t_0)q_0^2\biggr]+\frac{1}{2}\biggr\{\frac{1}{G(T)}\int_{t_0}^{t_f}dt[G(t-t_0)q_f+G(t_f-t)q_0]J(t)\\&-\frac{1}{G(T)}\int_{t_0}^{t_f}dt\int_{t_0}^{t_f}dt'~[\Theta(t-t')G(t_f-t)G(t'-t_0)+\Theta(t'-t)G(t_f-t')G(t-t_0)]J(t)J(t')\biggr\}.\end{split}
\ee

\section{\textcolor{blue}{\textbf{\large Generating function}}}\label{genfuncsk}
\begin{widetext}
In this section, {the prime objective is to} derive the expression for the generating function {for the time dependent} inverted oscillator,
using the action that we computed in the previous section.\\
The generating function for a particular classical action is given by,
\be\label{genZ}
\displaystyle
Z(q_0,t_0;q_f,t_f|J)= C(T) \exp\left(iS_{cl}(q_0,t_0;q_f|J)\right),
\ee
where {$T=t_{f}-t_{0}$} and $C(T)$ is an overall factor that does not depend on the external source $J$. Hence we can rewrite,
\be\label{defZ}
Z(q_0,t_0;q_f,t_f):=
Z(q_0,t_0;q_f,t_f|0)=C(T) \exp\left(iS_{cl}(q_0,t_0;q_f,t_f|0)\right).
\ee
\end{widetext}
The generating function without the source i.e. $Z(q_0,t_0;q_f,t_f)$ can be represented by transition amplitude $\bra{q_f}e^{i\hat{H}T}\ket{q_0}$ and can even be decomposed as shown below :
\begin{widetext}
\be\label{fgf}
\begin{aligned} Z(q_0,t_0;q_2,t_2)=\bra{q_2}e^{-it_2\hat{H}}e^{it_0\hat{H}}\ket{q_0} =\bra{q_2}e^{-it_2\hat{H}}e^{it_1\hat{H}}e^{-it_1\hat{H}}e^{it_0\hat{H}}\ket{q_0}~~~~~~~~~~~~~~~~~~~~~~~~~~~~~~~~\\=\int_{-\infty}^{\infty}dq_1\bra{q_2}e^{-it_2\hat{H}}e^{it_1\hat{H}}\ket{q_1}\bra{q_1}e^{-it_1\hat{H}}e^{it_0\hat{H}}\ket{q_0}~~~~~~~~~\\=\int_{-\infty}^{\infty}dq_1 Z(q_1,t_1;q_2t_2|0) Z(q_0,t_0;q_1,t_1|0) ~~~~~~~~~~~~~~~~~~~~
\end{aligned}
\ee
Applying the composition law shown in  Eq. (\ref{fgf}) we can rewrite Eq.\eqref{genZ} as:
\be\label{Zcomp}
\int_{-\infty}^{\infty}dq_1 Z(q_1,t_1;q_2t_2|0) Z(q_0,t_0;q_1,t_1|0)=C(t_1-t_0)C(t_2-t_1){e^{i(S_{cl}(q_0,t_0;q_1,t_1)+S_{cl}(q_1,t_1;q_2,t_2))}}
\ee
From Eq.\eqref{action} we can write the action as,
  \be \label{action0} S_{cl}(q_0,t_0;q_f,t_f|0)= \frac{\dot{G}(t_f-t_0)\biggr[(m(t_f)q_f^2+m(t_0)q_0^2\biggr]-\biggr[m(t_f)+m(t_0)\biggr]q_fq_0}{2G(t_f-t_0)} -\frac{1}{2} \biggr[f(t_f)m(t_f)q_f^2-f(t_0)m(t_0)q_0^2\biggr].
\ee
If one considers that the system undergoes quantum dynamical event from $q_0$  to $q_1$ and $q_1$ to $q_2$, one can use Eq.\eqref{action0} to find the sum of actions {obtained from} the classical solution. It can then be shown that,
\end{widetext}
\begin{widetext}
\begin{flalign}\label{sum1}
&S_{cl}(q_{0},t_{0};q_{1},t_{1}|0)+S_{cl}(q_{1},t_{1};q_{2},t_{2}|0)=\frac{N}{2G(t_1-t_0)G(t_2-t_1)} -\frac{1}{2}\biggr[f(t_2)m(t_2)q_2^2-f(t_0)m(t_0)q_0^2\biggr],
\end{flalign}
where, {the normalization factor N is given by the following expression:}
\begin{flalign}
\nonumber N=G(t_2-t_0)q_1^2+G(t_2-t_1)&\dot{G}(t_1-t_0)m(t_0)q_0^2+G(t_1-t_0)\dot{G}(t_2-t_0)m(t_2)q_2^2
\\&-q_1\biggr\{[m(t_1)+m(t_0)]q_0G(t_2-t_1)+[m(t_1)+m(t_2]q_2G(t_1-t_0)\biggr\}.
\end{flalign}
Using algebraic manipulation one can rewrite {the expression for the normalization factor N as:}
\begin{flalign}\label{ufz}
   \nonumber N=G(t_2-t_0)m(t_1)&\biggr[q_1-\frac{2q_1[m(t_1)+m(t_0)]G(t_2-t_1)q_0+[m(t_2)+m(t_1)]q_2G(t_1-t_0)}{2m(t_1)G(t_2-t_0)}\biggr]^{2}+G(t_2-t_1)\dot{G}(t_1-t_0)m(t_0)q_0^2\\&+G(t_1-t_0)\dot{G}(t_2-t_1)m(t_2)q_2^2%\displaystyle\\
-\frac{\biggr[[m(t_1)+m(t_0)]G(t_2-t_1)q_0+[m(t_2)+m(t_1)G(t_1-t_0)\biggr]^2}{2m(t_1)G(t_2-t_0)}&&.
%\end{matrix}
\end{flalign}
Note that to avoid long equations henceforth we denote:\\  $$\biggr[q_1-\frac{2q_1[m(t_1)+m(t_0)]G(t_2-t_1)q_0+[m(t_2)+m(t_1)]q_2G(t_1-t_0)}{2m(t_1)G(t_2-t_0)}\biggr]^{2}\equiv\biggr(q_1\dots\biggr)^2$$

To find the generating function using Eq. (\ref{Zcomp}) one needs to evaluate integration over $q_1$. We therefore rearrange the terms in above equation as follows:\\
\be 
\begin{aligned}\label{nile}
N-G(t_2-t_1)m(t_1)(q_1\dots)^2=~~~~~~~~~~~~~~~~~~~~~~~~~~~~~~~~~~~~~~~~~~~~~~~~~~~~~~~~~~~~~~~~~~~~~~~~~~~~~~~~~~~~~~~~~~~~~~~~~~~~~~~~~~~~~~~~~~~~~~~~~~~~~~~~~~~~~\\\frac{1}{{2m(t_1)G(t_2-t_0)}}\biggr(2m(t_0)m(t_1)G(t_2-t_0)G(t_2-t_1)\dot{G}(t_1-t_0)q_0^2+2m(t_1)m(t_2)G(t_1-t_0)\dot{G}(t_2-t_1) G(t_2-t_0)q_2^2~~~~~~~~~~~~~~~~~~~~~~~~~~~\\~~~~~~~-m^2(t_1)G^2(t_2-t_1)q_0^2-2m(t_1)m(t_0)G^2(t_2-t_1)q_0^2-m^2(t_0)G(t_2-t_1)q_0^2-m^2(t_2)G^2(t_1-t_2)q_2^2~~~~~~~~~~~~~~~~~~~~~~~~~~~~\\-2m(t_2)m(t_1) G^2(t_1-t_0)q_2^2-m^2(t_1)G^2(t_1-t_0)q_2^2-2q_0q_2G(t_1-t_0)G(t_2-t_1)[m(t_1)+m(t_0)][m(t_2)+m(t_1)]\biggr).~~~~~~~~~~~~~~~~~~~~~
\end{aligned}
\ee
The hyperbolic relations in terms of Green's function and its time derivatives are given below,
\be \label{evalue}
\begin{aligned}
G(t_2-t_0)\dot{G}(t_1-t_0)-G(t_2-t_1)=\dot{G}(t_2-t_0)G(t_1-t_0)\\ G(t_2-t_0)\dot{G}(t_2-t_1)-G(t_1-t_0)=\dot{G}(t_2-t_0)G(t_2-t_1).
\end{aligned}
\ee
Using these hyperbolic relations one can {further} simplify the Eq \eqref{nile} and write {it in the present form:}\\
\be
\begin{aligned}
N-G(t_2-t_1)m(t_1)\biggr(q_1\dots\biggr)^2=\frac{1}{2m(t_1)G(t_2-t_0)}\Big\{2m(t_0)m(t_1)G(t_2-t_1)\dot{G}(t_2-t_0)G(t_1-t_0)q_0^2~~~~~~~~~~~~~~~~~~~~~~~~~~~~~~~~~~~~~~~~~~~~~~~~~~~~~~~~~~~~~~~~~~~~~~~~~~~~~~~\\+2m(t_1)m(t_2)G(t_1-t_0)\dot{G}(t_2-t_0)G(t_2-t_1)q_2^2-[m^2(t_1)+m^2(t_0)]G^2(t_2-t_1)q_0^2~~~~~~~~~~~~~~~~~~~~~~~~~~~~~~~~~~~~~~~~~~~~~~~~~~~~~~~~~~\\-[m^2(t_1)+m^2(t_2)] G^2(t_1-t_0)q_2^2-2q_0q_2G(t_1-t_0)G(t_2-t_1)[m(t_1)+m(t_0)][m(t_2)+m(t_1)]\Big\}.~~~~~~~~~~~~~~~~~~~~~~~~~~~~~~~~~~~~~~~~~~~~~~~~~~~
\end{aligned}
\ee
Using the above expression, the sum of classical actions in Eq.\eqref{sum1} can be simplified to the {following} form,
\be
\begin{aligned}\displaystyle
S_{cl}(q_{0},t_{0};q_{1},t_{1}|0)+S_{cl}(q_{1},t_{1};q_{2},t_{2}|0)=~~~~~~~~~~~~~~~~~~~~~~~~~~~~~~~~~~~~~~~~~~~~~~~~~~~~~~~~~~~~~~~~~~~~~~~~~~~~~~~~~~~~~~~~~~~~~~~~~~\\S_{cl}(q_0,t_0;q_2,t_2|0)+\frac{G(t_2-t_0)m(t_1)}{2G(t_1-t_0)G(t_2-t_1)}\biggr[q_1-\dots \biggr]^2\displaystyle+\frac{q_2q_0}{2G(t_2-t_0)}\biggr[m(t_1)\frac{m(t_0)m(t_2)}{m(t_1)}\biggr]\\-\frac{\biggr\{\biggr[\frac{m^2(t_1)+m^2(t_0)}{m(t_1)}\biggr]G^2(t_2-t_1)q_0^2+\biggr[\frac{m^2(t_1)+m^2(t_2)}{m(t_1)}\biggr]G^2(t_1-t_0)q_2^2\biggr\}}{4G(t_2-t_1)G(t_1-t_0)G(t_2-t_0)}.\\\end{aligned}
\ee
%read from here%
The generating function of Eq.\eqref{Zcomp} can then be evaluated as: \\
\be \label{Zf}\begin{aligned}
Z(q_2,t_2;q_0,t_0|0)=C(t_1-t_0)C(t_2-t_1)e^{iS_{cl}(q_0,t_0;q_2,t_2|0)} \exp{(\mathcal{R})} \displaystyle\int_{-\infty}^{\infty}dq_1\exp{\biggl[\frac{iG(t_2-t_0)m(t_1)}{2G(t_1-t_0)G(t_2-t_1)}\biggr\{q_1^2-\dots\biggr\}\biggr]}.~~~~~~~~~~~~~~~~~~~~~~~~~~~~~~~~~~~~~~~~~~~~~~~~\\\end{aligned}
\ee
{Here, the factor $\mathcal{R}$ in the exponent is given by:}
\be
\mathcal{R}
=\Bigg[i\biggr\{\frac{q_0q_2\biggr[m(t_1)-\frac{m(t_0)m(t_1)}{m(t_1)}\biggr]}{2G(t_2-t_0)}-\frac{\biggr\{\biggr[\frac{m^2(t_1)+m^2(t_0)}{m(t_1)}\biggr]G^2(t_2-t_1)q_0^2+\biggr[\frac{m^2(t_1)+m^2(t_2}{m(t_1)}\biggr]G^2(t_1-t_0)q_2^2\biggr\}}{4G(t_2-t_1)G(t_1-t_0)G(t_2-t_0)}\biggr\}\Bigg].
\ee
Solving the integral over $q_1$ in Eq.\eqref{Zf} we obtain:
\bea
Z(q_2,t_2;q_0,t_0|0)=C(t_1-t_0)C(t_2-t_1)e^{iS_{cl}(q_2,t_2;q_0,t_0|0)}\exp{(\mathcal{R})}\sqrt{\frac{2\pi iG(t_2-t_1)G(t_1-t_2)}{G(t_2-t_0)m(t_1)}}.
\eea
Using Eq.\eqref{defZ} one can evaluate the LHS of the above equation and hence, we write:
\bea \nonumber
C(t_2-t_0)e^{iS_{cl}(q_2,t_2;q_0,t_0|0)}=C(t_1-t_0)C(t_2-t_1)e^{iS_{cl}(q_2,t_2;q_0,t_0|0)}\exp{(\mathcal{R})}\sqrt{\frac{2\pi iG(t_2-t_1)G(t_1-t_2)}{G(t_2-t_0)m(t_1)}}.\\
\eea
From the above equation, we can easily find:
\bea\label{Coft}
C(T)=
\exp{(-\mathcal{R})}\sqrt{\frac{{m(t_1)}}{{2\pi iG(T)}}}
\eea
where, $G(T)=G(t_{f}-t_{0})$ is the Green's function.
{Finally}, substituting Eq.\eqref{Coft} in Eq.\eqref{Zf} the final expression for the generating function of inverted oscillator is,
\begin{equation}    
\begin{aligned}
Z(q_{2},t_{2};q_{0},t_{0}|0)=\frac{\sqrt{m(t_{1})}}{\sqrt{2\pi iG(T)}}&\exp\Biggl[i\biggl\{-q_{1}q_{2}\frac{\biggl\{m(t_{1})-\frac{m(t_{0})m(t_{2})}{m(t_{1})}\biggr\}}{2G(t_{2}-t_{0})}\\+&\frac{\biggr\{\biggl[\frac{m^{2}(t_{1})+m^{2}(t_{0})}{m(t_{1})}\biggr]G^{2}(t_{2}-t_{1})q_{0}^{2}+\biggl[\frac{m^{2}(t_{1})+m^{2}(t_{2})}{m(t_{1})}\biggr]G^{2}(t_{1}-t_{0})q_{2}^{2}\biggr\}}{4G(t_{2}-t_{1})G(t_{1}-t_{0})G(t_{2}-t_{0})}\biggr\}\Biggr] e^{i S_{cl}(q_{2},t_{2};q_{0},t_{0}|0)}.~~~~~~~~~~~~~~~~~~~~~~~~
\end{aligned}
\end{equation}
   
    %\be
    %\begin{aligned}
    
    %\end{aligned}
    %\ee
\end{widetext}
\section{\textcolor{blue}{\textbf{\large Schwinger-Keldysh Path Integral}}}\label{SKpath}
\begin{figure}
    \centering
   \includegraphics[width=15cm, height=8cm]{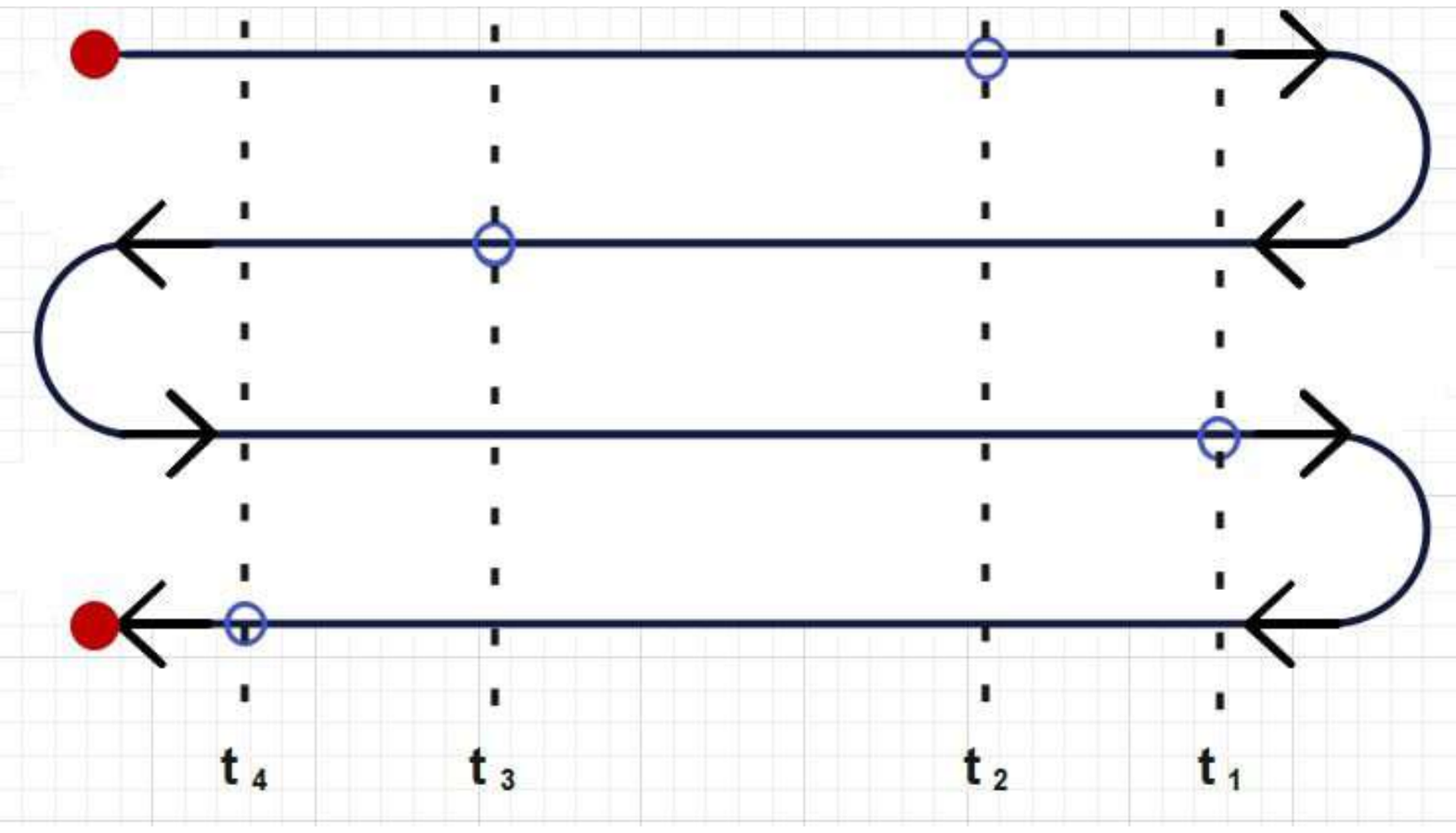}
   \caption{The schematic representation of Schwinger-Keldysh contour for the correlator functions with temporal ordering}
    \label{fig:my_label}
\end{figure}

In this section we give a brief idea of how Feynman path-integral leads to the Schwinger-Keldysh path Integral for non-equlibrium systems. Moreover we study in detail the generating function using the Schwinger-Keldysh path integral for inverted oscillator.  Note that we follow the prescription given in ref. \cite{BenTov:2021jsf} for an easy comparison.\\
The propagation function or generating function of Feynman path integral can be expressed as:
\be\label{feynman}
Z(q_0,t_0;q,t|0)=\int_{q(t_0)=q_0}^{q(t)=q}\mathcal{D}q(\cdot)~e^{\int_{t_0}^{t}\mathcal{L}(q(t'))~dt'}.
\ee
~~~In Feynman's path integral formalism, one can calculate the moments of distribution by introducing an auxiliary variable $(J)$ and taking the derivative of the generating function with respect to the $J$.
The one point amplitude is the given by:
 \be\label{moment}
 \bra{q_{f}}e^{-i\hat{H}t_{f}}\hat{q}(t_{1})e^{i\hat{H}t_{0}}\ket{q_{0}}=-i\frac{\delta}{\delta J(t_{1})}Z(q_{0},q_{f}|J)\biggr|_{J=0}.
 \ee
For multiple field configurations, $Z(q_{0},q_{f}|J)$ could generate the amplitude of time-ordered products of fields:
\be
\bra{q_{f}}e^{-i\hat{H}t_{f}}\mathcal{T}(\hat{q}(t_{1})\dots \hat{q}(t_{n}) e^{i\hat{H}t_{0}}\ket{q_{0}}=(-i)^{n}\frac{\delta^{n}}{\delta J(t_{1})\dots\delta J(t_{n})}Z(q_{0},q_{f}|J)\biggr|_{J=0}.\ee
Here $\mathcal{T}$ represents the time ordering symbol. For out-of-equlibrium systems like the inverted oscillator, although we need to compute such correlators, we cannot use time-ordering and hence must use some other formalism.  The Schwinger-Keldysh contour facilitates refining to a time-folded contour and hence provide a path-integral representation of the out-of-time ordered correlators. In Fig.\ref{fig:my_label} the four legs are denoted by each time stamp $t_{1}, t_{2}, t_{3}, t_{4}$ and we have taken the past-turning point in the contour at time $t_{1}$ and future-turning point at certain time $t$. Now, for action of an operator say $\theta$ at $t_{0}$, at a later time $t_{1}>t_{0}$, it must be related by the usual Heisenberg evolution with unitary time evolution operator $U(t, t_{0})$ i.e. 
 \begin{equation}
     \theta(t)=U(t, t_{0})\theta(t_{0})U^{\dagger}(t, t_{0}).
 \end{equation}
 
The source-free Feynman path-integral is nothing but an evolution operator and it is related to the Schrodinger picture wave function, if $\psi(q)=\langle q|\psi\rangle$. So we can write:
\be
\Psi(q,t)=\int_{-\infty}^{\infty}dq_0\langle q|\hat{U}(t-t_0)|q_0\rangle \langle q_0|\psi\rangle=\int_{-\infty}^{\infty}dq_0~\psi(q_0)~\int_{q_0}^{q_t}~\mathcal{D}q(\cdot)e^{\int_{t_0}^{t}~dt'~\mathcal{L}(q(t'))}.
\ee
Note that, the transportation event can be written as $\bra{q}U(t-t_0)\ket{q_0}$, this mathematical expression tells us the history of propagation event from $q_0$ to $q$.

When we consider a special case that  the wave density matrix $\hat{\rho}=\ket{q_0}\bra{q_0}$(Ground state) is a pure state with definite coordinates, then the one-point correlator can be written as,
\be\label{amplitude}\mathcal{T}(\hat{q}(t_1)~
e^{i\hat{H}t_0}~\hat{\rho}~e^{-i\hat{H}t_0})=
\bra{q_0}e^{-i\hat{H}t_0}\hat{q}(t_1)~e^{i\hat{H}t_0}\ket{q_0}.
\ee

On rewriting the above equation by introducing the identity operator $\hat{\rho}$ we can write the one point correlator as:
\be
\mathcal{T}(\hat{q}(t_1)~e^{i\hat{H}t_0}~\hat{\rho}~e^{-i\hat{H}t_0})=\int_{-\infty}^{\infty}dq_f\int_{-\infty}^{\infty}dq_0\int_{-\infty}^{\infty}dq'_0\bra{q_f}e^{-i\hat{H}t_f}\hat{q}(t_1)~e^{i\hat{H}t_0}\ket{q_0}\bra{q_0}\hat{\rho}\ket{q'_0}\bra{q'_0}e^{-i\hat{H}t_0}~e^{i\hat{H}t_f}\ket{q_f}.
\ee
Using Eq.\eqref{moment} one can then write:
\be
\mathcal{T}(\hat{q}(t_1)~e^{i\hat{H}t_0}~\hat{\rho}~e^{-i\hat{H}t_0})=(-i)\frac{\delta}{\delta J(t_1)}\int_{-\infty}^{\infty}dq_f\int_{-\infty}^{\infty}dq_0\int_{-\infty}^{\infty}dq'_0~\rho(q_0,q'_0)~Z(q_0,q_f|J)Z(q'_0,q_f|J')^*\biggr|_{J=J'=0}.\\\ee
Hence, we can define the generating function corresponding to the Schwinger-Keldysh Path-integral, by comparing the above equation with Eq.\eqref{amplitude} as:
\be\label{skgf}
Z(J,J')=\int_{-\infty}^{\infty}dq_f\int_{-\infty}^{\infty}dq_0\int_{-\infty}^{\infty}dq'_0~\rho(q_0,q'_0)~Z(q_0,q_f|J)Z(q'_0,q_f|J')^*.
\ee
In Eq. (\ref{skgf}), if the state is pure i.e. $\hat{\rho}=\ket{\psi}\bra{\psi}$, then we can factorise the double integral over $q_0$, that is initial field configuration, as shown below:
\be\label{rumination}\begin{aligned}
\displaystyle\int_{-\infty}^{\infty}dq_0\int_{-\infty}^{\infty}dq'_0~\rho(q_0,q'_0)~Z(q_0,q_f|J)Z(q'_0,q_f|J')^*~~~~~~~~~~~~~~~~~~~~~~~~~~~~~~~~~~~~~~~~~~~~~~~~~~~~~~~~~~~~\\\displaystyle=\biggr(\int_{-\infty}^{\infty}dq_0~\langle q_0 | \psi \rangle Z(q_0,q_f|J)\biggr)\biggr(\int_{-\infty}^{\infty}dq'_0~\langle q'_0 | \psi \rangle~Z(q'_0,q_0|J')\biggr)^*\\
\end{aligned}
\ee
On can rewrite the above equation as,
\be
\int_{-\infty}^{\infty}dq_0\int_{-\infty}^{\infty}dq'_0~\rho(q_0,q'_0)~Z(q_0,q_f|J)Z(q'_0,q_f|J')^*=\Psi(q_f|J)\Psi(q_f|J')^*,
\ee
where, {we define:}
\be
\Psi(q_f|J)=\int_{-\infty}^{\infty}~dq_0~\langle q_0 |0\rangle ~Z(q_0,q_f|J).
\ee
Now, by using the wave function \eqref{psiuf} for the ground state of inverted oscillator given by,
\be
\langle q | 0\rangle=\frac{\sqrt{a}}{(2\pi)^\frac{1}{4}}\biggr(\frac{1}{2\pi K(t)}\biggr)^{\frac{1}{2}} e^{\frac{-q^2}{2x_0^2}}
\ee
and the density matrix  $\hat{\rho}=\ket{0}\bra{0}$, we can express the generating function in its factorised form:
\be\label{gfn}
Z(J,J')=\int_{-\infty}^{\infty}dq_f~\Psi(q_f|J)\Psi(q_f|J')^*.
\ee
Next, by using the equation for classical solution for the action given in the Eq. (\ref{action})  we can write,\\
\be
\begin{aligned}
\Psi(q_f|J)=\frac{\sqrt{a}}{(2\pi)^\frac{1}{4}}\biggl(\frac{1}{2\pi K(t)}\biggr)^\frac{1}{2}\frac{\sqrt{m(t_1)}}{\sqrt{2\pi G(T)}}\exp\Bigg[\frac{i}{2G(T)}\biggl[m(t_f)\dot{G}(T)q_f^2-G(T)f(t_f)m(t_f)q_f^2~~~~~~~~~~~~~~~~~~~~~~~~~~~~~~\\+\frac{\biggr[\frac{m^2(t_1)+m^2(t_f)}{m(t_1)}\biggr]G(t_1-t_0)q_f^2}{2G(t_f-t_1)}+[m(t_f)+1]\int_{t_0}^{t_f}dt' q_f G(t'-t_0)J(t')\\-2\int_{t_0}^{t_f}dt\int_{t_0}^{t_f}dt'\Theta(t-t')G(t_f-t')G(t'-t_0)\biggr]J(t)J(t')\Bigg]\chi(q_f|J).
\end{aligned}\label{psif}
\ee
Here $\chi(q_f|J)$ denotes the contribution of integration over initial field configuration i.e. $q_0$. We compute this integral's contribution in the Appendix \ref{intif}. Inserting Eq.\eqref{xif} in Eq.\eqref{psif} one can show that,
\begin{equation}\label{Psi}
\begin{aligned}   %\displaystyle
\Psi(q_f|J)=\frac{\sqrt{a}}{(2\pi)^{\frac{1}{4}}}\biggr(\frac{1}{2\pi K(t)}\biggr)^\frac{1}{2}\frac{\sqrt{m(t_1)}}{\sqrt{2\pi G(T)}}\sqrt{2\pi iG(T)\mathscr{A}^{-1}(t)}\exp\Bigg[\frac{i}{2G(T)}\biggr\{m(t_f)\dot{G}(T)q_f^2-G(T)f(t_f)m(t_f)q_{f}^{2}~~~~~~~~~~~~~~~~~~~\\+\frac{\biggr[\frac{m^{2}(t_1)+m^{2}(t_f)}{m(t_1)}\biggr]G(t_1-t_0)q_{f}^{2}}{2G(t_f-t_1)}+[m(t_f)+1]\int_{t_0}^{t_f}dt'~q_fG(t_1-t_0)J(t')~~~~~~~~~~~~~~~~~~~~~~~~~~~~~~~~~~~~~~~~~~~~~~~~~~~~\\-2\int_{t_0}^{t_f}dt\int_{t_0}^{t_f}dt'[\Theta(t-t')G(t_f-t')G(t'-t_0)]J(t)J(t')\bigg\}\Bigg]\exp\biggr[\frac{-i\mathscr{A}^{-1}(t)}{8G(T)}\times ~~~~~~~~~~~~~~~~~~~~~~~~~~~~~~~~~~~~~~~~~\\\biggr[\biggr\{m(t_f)+m(t_0)+m(t_1)-\frac{m(t_0)m(t_f)}{m(t_1)}\biggr\}q_{f}-[m(t_0)+1]\int_{t_0}^{t_f}dt~G(t_f-t)J(t)\biggr]^2\biggr].~~~~~~~~~~~~~~~~~~~~~~~~~~~~~~~~~~~~~
\end{aligned}
\end{equation}
By performing similar steps one can find $\Psi^{*}(q_f|J')$.

Substituting Eq.\eqref{Psi} in Eq.\eqref{gfn} one can write the generating function as:
\be\label{zjj}
\begin{aligned}
Z(J,J')=m(t_1)\frac{a}{\sqrt{2\pi}}\frac{\mathscr{A}^{-1}(t)}{2\pi K(t)}\exp\biggr[\frac{i}{2G(T)}[m(t_f)+1]\int_{t_0}^{t_f}dt'~q_fG(t_1-t_0) [J(t')-J'(t')]~~~~~~~~~~~~~~~~~~~~~~~~~~~~~~~~~~~~~~~~~~~~~~~~~~~~~~~~~~~~~~~~~~~~~~~~~~~~~~~~~~\\~+2\int_{t_0}^{t_f}dt\int_{t_0}^{t_f}dt'[\Theta(t-t')G(t_f-t')G(t'-t_0)][J'(t)J'(t')-J(t)J(t')]\biggr\}\biggr]\int_{-\infty}^{\infty}dq_{f}\exp\biggr\{\mathcal{I}(J,J')\biggr\}.~~~~~~~~~~~~~~~~~~~~~~~~~~~~~~~~~~~~~~~~~~~~~~~~\\
\end{aligned}
\ee
Here $\mathcal{I}(J,J')$ contains the terms having contribution of the final field configuration, $q_f$ given by:
\be
\begin{aligned} \label{Ijj}
\mathcal{I}(J,J')=\biggl[\frac{i\mathscr{A}^{-1}(t)}{8G(T)}\biggl(\biggl\{m(t_f)+m(t_0)+[m(t_1)-\frac{m(t_0)m(t_f)}{m(t_1)}\biggr\}q_f-[m(t_0)+1]\int_{t_0}^{t_f}dt~G(t_f-t)J(t)\biggr)^2\biggr]~~~~~~~~~\\ +\biggl[\frac{-i\mathscr{A}^{-1}(t)}{8G(T)}\biggl(\biggl\{m(t_f)+m(t_0)+[m(t_1)-\frac{m(t_0)m(t_f)}{m(t_1)}\biggr\}q_f-[m(t_0)+1]\int_{t_0}^{t_f}dt~G(t_f-t)J(t)\biggr)^{2}\biggr].
\end{aligned}
\ee

Since, $\mathscr{A}^{-1}(t)$ as given in Eq.\eqref{at}, is real, the $q_{f}^{2}$ terms in Eq.\eqref{Ijj} will be cancelled out and the only term that contains $q_{f}$ will be considered for the integration over the final configuration. The expression of $\mathcal{I}(J,J')$ upon simplification is then given by:
\be
\begin{aligned}\label{particu}
\mathcal{I}(J,J')=\frac{4i\mathscr{A}^{-1}(t)}{8G(T)}\Biggl[-q_{f}\frac{[m(t_0)+1]}{2}\biggl[m(t_f)+m(t_0)+m(t_1)-\frac{m(t_0)m(t_f)}{m(t_1)}\biggr]\biggl\{\int_{t_0}^{t_f}dt'G(t_f-t')[J'(t')-J(t)]\biggr\}\displaystyle\\-\biggl[\frac{[m(t_0)+1]}{2}\int_{t_0}^{t_f}dt'G(t_f-t')J(t')\biggr]^2+\biggl[\frac{[m(t_0)+1]}{2}\int_{t_0}^{t_f}dt'G(t_f-t')J'(t')\biggr]^2 \Biggr]~~~~~~~~~~~~~~~~~~~~~~~~\\
\end{aligned}
\ee
We write a particular term of Eq.\eqref{particu} as:
\bea \label{manipulation}
\mathscr{A}^{-1}(t)\biggr[\frac{[m(t_0)+1]}{2}\int_{t_0}^{t_f}dt'G(t_f-t')J(t')\biggr]^2=\mathscr{A}^{-1}(t)\frac{[m(t_0)+1]^2}{4}\int_{t_0}^{t_f}dt\int_{t_0}^{t_f}dt'G(t_f-t')G(t_f-t)J(t)J(t').~~~~~\
\eea
%\subsection{\textcolor{Sepia}{\textbf{\normalsize Integration over final field configuration}}}
Now, putting Eq.\eqref{manipulation} back in Eq \eqref{particu} and then using the final expression of $\mathcal{I}(J,J')$ in Eq.\eqref{zjj} we can write the generating function for inverted oscillator as:
\be\label{generating }
\begin{aligned}
Z(J,J')=m(t_1)\frac{a}{\sqrt{2\pi}}\frac{1}{2\pi K(t)}\frac{1}{|\mathscr{A}(t)|}\exp\biggr[\frac{i}{G(T)}\int_{t_0}^{t_f}dt\int_{t_0}^{t_f}dt'[\Theta(t-t')G(t_f-t)G(t'-t_0)][J'(t)J'(t')-J(t)J(t')]\biggr]~~~~~~~~~~~~~~~~~~~~~~~~~~~~~~~~~~~~~~~~~~~~\\\times\int_{-\infty}^{\infty}dq_f\exp\Bigg[\frac{i\mathscr{A}^{-1}(t)}{2G(T)}\Biggr\{\frac{[m(t_0)+1]}{2}\bigg[m(t_f)+m(t_0)+m(t_1)-\frac{m(t_0)m(t_f)}{m(t_1)}\bigg]\bigg\{\int_{t_0}^{t_f}dt'G(t_f-t')[J(t')-J'(t)]\bigg\}q_f~~~~~~~~~~~~~~~~~~~~~~~~~~~~~~~~~~~~~~~~~~~~\\+\frac{[m(t_0)+1]^2}{4}\biggr[\int_{t_0}^{t_f}dt\int_{t_0}^{t_f}dt'G(t_f-t')G(t_f-t)[J'(t)J'(t')-J(t)J(t')]\biggr]+[m(t_{f})+1]\mathscr{A}(t)\int_{t_0}^{t_f}dt'q_fG(t'-t_0)\times ~~~~~~~~~~~~~~~~~~~~~~~~~~~~~~~~~~~~~~~~~~~~\\~[J(t')-J'(t')]\Bigg\}\Biggr].~~~~~~~~~~~~~~~~~~~~~~~~~~~~~~~~~~~~~~~~~~~~~~~~~~~~~~~~~~~~~~~~~~~~~~~~~~~~~~~~~~~~~~~~~~~~~~~~~~~~~~~~~~~~~~~~~~~~~~~~~~~~~~~~~~~~~~~~~~~~~~~~~~~~~~~~~~~~~~~~~~~~~~~~~~\\
\end{aligned}
\ee
%%%%%%
Next we define,
\begin{flalign}
j&\nonumber=\frac{[m(t_{0})+1]}{2}\bigg[m(t_f)+m(t_0)+m(t_1)-\frac{m(t_0)m(t_f)}{m(t_1)}\bigg]\int_{t_0}^{t_f}dt'~\biggr[G(t_f-t')[J'(t')-J(t)]\biggr]\\
&~~~~+[m(t_{f}+1]\mathscr{A}(t)\int_{t_0}^{t_f}dt'~G(t'-t_0)[J(t')-J'(t')].
\end{flalign}
%%%%%%
Then the integral over final field configuration in Eq.\eqref{generating } can be evaluated as:
\be
\int_{-\infty}^{\infty}\exp{\biggr(\frac{i\mathscr{A}^{-1}(t)}{2G(t)}jq_{f}}\biggr) dq_{f} =4 \pi G(T)\mathscr{A}(t) \delta(j).
\ee.

Finally, the generating function can be written as,
\be\label{last}
\begin{aligned}
Z(J,J')=\frac{2m(t_{1})aG(T)}{\sqrt{2\pi} K(t)}~\delta
\biggr(\int_{t_0}^{t_f}dt'[m(t_{f})+1]\mathscr{A}(t)G(t'-t_0)+\frac{[m(t_{0})+1]}{2}\biggl[m(t_f)+m(t_0)+m(t_1)-\frac{m(t_0)m(t_f)}{m(t_1)}\biggr]~~~~~~~~~~~~~~~~~~~~~~~~~~~~~~~~~~~~~~~~~~~~~~~~~~~~~~~~~~~~~~~~~~~~~~~~~~~~~~~~~~~~~~~~~~~~~~~~~~~~~~~~~~~~~~~~~~~~~~~~~~~~~~~~~~~~~~~~~~~~~~~~~~~~~~~~~~~~~~~~~~~~~~~~~~~~~~~~~~~~~~~~~~~~\\\times G(t_f-t')[J(t')-J'(t)]\biggr) \exp\Biggr[\frac{i}{2G(T)}\biggr\{\frac{[m(t_0)+1]^2}{4\mathscr{A}(t)}\biggr[\int_{t_0}^{t_f}dt\int_{t_0}^{t_f}dt'G(t_f-t')G(t_f-t)[J'(t)J'(t')~~~~~~~~~~~~~~~~~~~~~~~~~~~~~~~~~~~~~~~~~~~~~~~~~~~~~~~~~~~~~~~~~~~~~~~~~~~~~~~~~~~~~~~~~~~~~~~~~~~~~~~~~~~~~~~~~~~~~~~~~~~~~~~~~~~~~~~~~~~~~~~~~~~~~~~~~~~~~~~~~~~~~~~~~~~~~~~~~~~~~~~~~~~~~~~~~~~~~~\\-J(t)J(t')]\biggr]+2\int_{t_0}^{t_f}dt\int_{t_0}^{t_f)}dt'[\Theta(t-t')G(t_f-t)G(t'-t_0)][J'(t)J'(t')-J(t)J(t')]\biggr\}\Biggr].~~~~~~~~~~~~~~~~~~~~~~~~~~~~~~~~~~~~~~~~~~~~~~~~~~~~~~~~~~~~~~~~~~~~~~~~~~~~~~~~~~~~~~~~~~~~~~~~~~~~~~~~~~~~~~~~~~~~~~~~~~~~~~~~~~~~~~~~~~~~~~~~~~~~~~~~~~~~~~~~~~~~~~~~~~~~~~~~~~~~~~~~~~~~~~~~~~~~~~~~~~~~~~~~~~~~~~
\end{aligned}
\ee
\subsection{\textcolor{blue}{\textbf{\normalsize Influence phase}}}
~~~In the case of Feynman path integral, the logarithm of the path integral is usually defined as "effective action." Similarly for the Schwinger-Keldysh Path-integral, the logarithm of the generating function of the path integral is defined by the "influence phase". The influence phase is given by,
\be
\Phi=-i\ln{[Z(J,J')]}.
\ee
Using Eq.\eqref{last}, the influence phase for inverted oscillator {is given by the following expression:}
\be
\begin{aligned}
\Phi=-i\ln\biggr[\frac{2m(t_{1})aG(T)}{\sqrt{2\pi} K(t)}\delta
\biggr(\int_{t_{0}}^{t_{f}}dt'[m(t_{f})+1]\mathscr{A}(t)G(t'-t_{0})+\frac{[m(t_{0})+1]}{2}\biggl[m(t_{f})+m(t_{0})+m(t_{1})-\frac{m(t_{0})m(t_{f})}{m(t_{1})}\biggr]~~~~~~~~~~~~~~~~~~~~~~~~~~~~~~~~~~~~~~~~~~~~~~~~~~~~~~~~~~~~~~~~~~~~~~~~~~~~~~~~~~~~~~~~~~~~~~~~~~~~~~~~~~~~~~~~~~~~~~~~~~~~~~~~~~~~~~~~~~~~~~~~~~~~~~~~~~~~~~~~~~~~~~~~~~~~~~~~~~~~~~~~~~~~~~~~~~~\\\times G(t_{f}-t')[J(t')-J'(t)]\biggr)\biggr]+\frac{1}{2G(T)}\biggr\{\frac{[m(t_{0})+1]^{2}}{4\mathscr{A}(t)}\biggr[\int_{t_{0}}^{t_{f}}dt\int_{t_{0}}^{t_{f}}dt'G(t_{f}-t')G(t_{f}-t)[J'(t)J'(t')-J(t)J(t')]\biggr]~~~~~~~~~~~~~~~~~~~~~~~~~~~~~~~~~~~~~~~~~~~~~~~~~~~~~~~~~~~~~~~~~~~~~~~~~~~~~~~~~~~~~~~~~~~~~~~~~~~~~~~~~~~~~~~~~~~~~~~~~~~~~~~~~~~~~~~~~~~~~~~~~~~~~~~~~~~~~~~~~~~~~~~~~~~~~~~~~~~~~~~~~~~~~~~~~\\+2\int_{t_{0}}^{t_{f}}dt\int_{t_{0}}^{t_{f}}dt'[\Theta(t-t')G(t_{f}-t)G(t'-t_{0})][J'(t)J'(t')-J(t)J(t')]\biggr\}.~~~~~~~~~~~~~~~~~~~~~~~~~~~~~~~~~~~~~~~~~~~~~~~~~~~~~~~~~~~~~~~~~~~~~~~~~~~~~~~~~~~~~~~~~~~~~~~~~~~~~~~~~~~~~~~~~~~~~~~~~~~~~~~~~~~~~~~~~~~~~~~~~~~~~~~~~~~~~~~~~~~~~~~~~~~~~~~~~~~~~~~~~~~~~~~~~~~~~~~~~~~~~~~~~~~~~~~~~~~~~~~~~~~~~~~~~~~~~~~~~~~~~
\end{aligned}
\ee

%==========================================

\section{\textcolor{blue}{\textbf{\large Calculation of Out-of-Time Order Correlator}}}\label{otocsec}
In this section we compute the Out-of-time ordered correlator (OTOC) as an arbitrarily ordered 4-point correlation function for the inverted oscillator using invariant operator method and Schwinger-Keldysh path integral formalism. Note that we closely follow the calculation of OTOC for harmonic oscillator \cite{BenTov:2021jsf} and modify the same for an inverted oscillator.  

To do so, the invariant operator given in Eq.\eqref{Inv} as:
\be\label{invariant op}
I(t)=-\frac{1}{2}\biggr[\biggr(\frac{\hat{q}}{K(t)}\biggr)^{2}-\biggr(K(t)\hat{P}-m(t)(\dot{K}(t)-f(t)K(t))\hat{q}^{2}\biggr)^{2}\biggr].
\ee
The invariant operator given above has the form of the forced harmonic oscillator which makes it hard to find the wave function $\psi(q,t)$ of the desired system. Because of this peculiar reason, it is necessary to perform the unitary transformation over the invariant operator. By doing so invariant operator method yields the straightforward calculation of the wave function.\\
~~~Now, the corresponding unitary operator can be represented as:
\be
\hat{U}=\exp{\biggr(-i\frac{m(t)}{K(t)}(\dot{K}(t)-f(t)K(t))\hat{q}^{2}\biggr)}.
\ee
If one performs the unitary transformation on the invariant operator of Eq.\eqref{invariant op}, then:
\begin{flalign}\label{Utranform}
    \bar{I}(t)&\nonumber=UI\bar{U}\\
    &\nonumber=\exp{\biggr(-\frac{im(t)}{K(t)}(\dot{K}(t)-f(t)K(t))\hat{q}^{2}\biggr)}\times\biggr[-\frac{1}{2}\biggr\{\biggr(\frac{\hat{q}}{K(t)}\biggr)^{2}-\biggr(K(t)\hat{P}-m(t)(\dot{K}(t)-f(t)K(t))\hat{q}^{2}\biggr)^{2}\biggr\}\biggr]\\
    &\nonumber\times\exp{\biggr(\frac{im(t)}{K(t)}(\dot{K}(t)-f(t)K(t))\hat{q}^{2}\biggr)}\\
    &\Rightarrow \bar{I}(t)=-\frac{1}{2}\biggr[\biggr(\frac{\hat{q}(t)}{K(t)}\biggr)^{2}-K^{2}(t)\hat{P}^{2}\biggr].
\end{flalign}
~~~The new eigenstate of the invariant operator after the diagonalization will be in the below given form,
\begin{flalign}
\varphi_{n}\rightarrow   \bar{\varphi}_{n}(\hat{q},t)&\nonumber=\hat{U}\varphi_{n}\\
   &=\exp{\biggr(-\frac{im(t)}{K(t)}(\dot{K}(t)-f(t)K(t))\hat{q}^{2}\biggr)}\varphi_{n}.
\end{flalign}
~~Then the eigenvalue equation for invariant operator becomes,
\begin{flalign}\label{de}
\nonumber\bar{I}(t)\psi_{n}(\hat{q},t)&=n\psi_{n}(\hat{q},t)\\
&\nonumber\Rightarrow -\frac{K^{2}(t)}{2}\frac{\partial^{2}\varphi_{(n)}}{\partial q^{2}}-\frac{1}{2K^{2}(t)}\hat{q}^{2}\varphi_{n}=n\varphi_{n}.\\
\end{flalign}
The solution to the Eq. \eqref{de} will be in terms of the Weber function:
\be
\varphi_{n}(\hat{q},t)=\frac{1}{\sqrt{K(t)}}\exp\biggl(\frac{im(t)}{2K(t)}(\dot{K}(t)-f(t)K(t))\hat{q}^{2}\biggr)\mathcal{D}_{n}\biggr(\sqrt{2}~\frac{\hat{q}^{2}(t)}{K(t)}\biggr)
\ee
~~~Where, $\mathcal{D}_{n}(x)$ is the Weber function.\\
Then the expression for the eigenstate of inverted oscillator in Eq.\eqref{psim} is modified in terms of Weber function as:
\begin{equation}
   \langle q| n\rangle=\Psi(\hat{q},t)=\frac{1}{\sqrt{K(t)}}\exp\biggl(-in\gamma_{n}(t)\biggr)\exp\biggl(\frac{im(t)}{2K(t)}(\dot{K}(t)-f(t)K(t))\hat{q}^{2}\biggr)\mathcal{D}_{n}\biggl(\sqrt{2}~\frac{\hat{q}^{2}(t)}{K(t)}\biggr).
\end{equation}
Here the phase factor is $\gamma_{n}(t)=\int_{0}^{t}\frac{dt'}{m(t')K^{2}(t')}$.

Using the form of the Green's function for inverted oscillator, computed in the Appendix \ref{greenf},  one can define the Heisenberg-picture field operator in terms of annihilation and creation operator:
\be\label{field op}
\hat{q}(t)=\frac{\Gamma_{0}}{\sqrt{2}}\biggr(\zeta^{*}(t)~a_{-}+\zeta(t)~a_{+}\biggr).
\ee
Here, $\zeta(t)=e^{\Gamma_{0}t}$~ and~ $\zeta^{*}(t)=e^{-\Gamma_{0}t}$. Note that $\Gamma_0$ is the factor used in Green's function for inverted oscillator given in Eq.\eqref{gfhyp}. We can obtain annihilation and creation operators for inverted oscillator  by putting $\omega(t)=i\omega(t)$ in the operators for any parametric oscillator \cite{Rajeev:2017uwk}:
\bea
\nonumber{a}_{-}&=\sqrt{\frac{i}{2}}\biggr[\biggr(\frac{\hat{q}}{K(t)}\biggr)+\biggr\{K(t)\hat{P}-m(t)[\dot{K}-f(t)K(t)]\hat{q}\biggr\}\biggr]\\
{a}_{+}&=\sqrt{\frac{i}{2}}\biggr[\biggr(\frac{\hat{q}}{K(t)}\biggr)-\biggr\{K(t)\hat{P}-m(t)[\dot{K}-f(t)K(t)]\hat{q}\biggr\}\biggr].
\eea
From this equation we can immediately argue that: ${a}^{\dagger}_{-}=i{a}_{-}; {a}^{\dagger}_{+}=i{a}_{+}$, and obviously $[{a}_{-},{a}_{+}]=1$ .\\
We can express the time evolution of the grounds state of the inverted oscillator using the Heisenberg field operator, $\hat{q}(t_{1})$, such that $t_{1}>t_{0}$:
\begin{flalign}\label{timeev}
  \ket{\psi(t_{1})}&\nonumber=\hat{q}(t_{1})~e^{it_{0}\hat{H}}\ket{0}\\
  &\nonumber=e^{it_{0}\hat{H}}e^{-it_{0}\hat{H}}\hat{q}(t_{1})~e^{it_{0}\hat{H}}\ket{0}\\
  &=e^{it_{0}\hat{H}}\hat{q}(t_{1}-t_{0})\ket{0}.
\end{flalign}
Similarly one can evolve the ground state from $t_1$ to $t_2$, using Heisenberg field operator, $\hat{q}(t_{2})$ as shown below,
\begin{flalign}\label{timeev2}
   \ket{\psi(t_{2})}&\nonumber=\hat{q}(t_{2})~e^{it_{0}\hat{H}}\ket{0}\\
  &\nonumber=e^{it_{1}\hat{H}}e^{-it_{1}\hat{H}}\hat{q}(t_{2})~e^{it_{0}\hat{H}}\ket{0}\\
  &=e^{it_{0}\hat{H}}\hat{q}(t_{2}-t_{1})\ket{0}.
\end{flalign}
Using the form of the Heisenberg field operator in Eq.\eqref{field op} the time evolution in Eq.\eqref{timeev} and Eq.\eqref{timeev2} can be expressed as,
\be\label{psitrstate}
  \ket{\psi(t_{1})}=e^{it_{0}\hat{H}}\frac{\Gamma_{0}}{\sqrt{2}}\zeta(t_{1}-t_{0})\ket{1},
\ee
\be\label{psitrstate1}
\ket{\psi(t_{2})}=e^{it_{0}\hat{H}}\frac{\Gamma_{0}}{\sqrt{2}}\zeta(t_{2}-t_{1})\ket{1}.
\ee
The two-point correlator for the ground state, $\hat{\rho}=\ket{0}\bra{0}$ of the inverted oscillator can be written as
\begin{flalign}\label{2p}
Tr\biggr(\hat{q}(t_{2})\hat{q}(t_{1})e^{it_{0}\hat{H}}~\hat{\rho}~e^{-it_{0}\hat{H}}\biggr) &\nonumber=\bra{0}e^{-it_{0}\hat{H}}\hat{q}(t_{2})~\hat{q}(t_{1})e^{it_{0}\hat{H}}\ket{0}\\
&=\langle \psi(t_{2}) |\psi(t_{1})\rangle.
\end{flalign}
Using Eq.\eqref{psitrstate} and Eq.\eqref{psitrstate1} the two point correlator for the ground state of the inverted oscillator becomes,
\be\label{2pointa}
Tr\biggr(\hat{q}(t_{2})\hat{q}(t_{1})e^{it_{0}\hat{H}}~\hat{\rho}~e^{-it_{0}\hat{H}}\biggr)=\frac{\Gamma^{2}_{0}}{2}\zeta(t_{1}-t_{0})~\zeta^{*}(t_{2}-t_{1}).
\ee
The two-point time evolution of the ground state of the inverted oscillator can be written as,
\be\label{4p}
\ket{\psi(t_{1},t_{2})}\equiv \hat{q}(t_{2})\hat{q}(t_{1})e^{i\hat{H}t_{0}}\ket{0}.
\ee
Using the unit operator $\hat{\textbf{1}}=e^{i\hat{H}t_{0}}e^{-i\hat{H}t_{0}}$, remembering that $t_{1,2}>t_{0}$, one can rewrite the above equation as:
\begin{flalign}\label{4point}
\ket{\psi(t_{1},t_{2})}&=e^{i\hat{H}t_{0}}e^{-i\hat{H}t_{0}} \hat{q}(t_{2})\hat{q}(t_{1})e^{i\hat{H}t_{0}}\ket{0}\\
&\nonumber=e^{i\hat{H}t_{0}}\hat{q}(t_{2}-t_{0})\hat{q}(t_{1}-t_{0})\ket{0}.
\end{flalign}
Inserting the form of the Heisenberg field operators, from Eq. \eqref{field  op} in Eq. \eqref{4point}, one can show that:\\
\begin{equation}\label{zetancon}
\ket{\psi(t_{1},t_{2})}=e^{i\hat{H}t_{0}}\frac{\Gamma_{0}}{\sqrt{2}}\zeta(t_{1}-t_{0})\biggr[\zeta^{*}(t_{2}-t_{0})a_{-}+\zeta(t_{2}-t_{0})a_{+}\biggr]\ket{1}.
\end{equation}
After operating the annihilation and creation operator on $\ket{1}$, Eq.\eqref{zetancon} becomes:
\begin{equation}\label{psizeta}
\ket{\psi(t_{1},t_{2})}=e^{i\hat{H}t_{0}}\frac{\Gamma_{0}}{\sqrt{2}}\zeta(t_{1}-t_{0})\biggr[\zeta^{*}(t_{2}-t_{0})\ket{0}+\sqrt{2}\zeta(t_{2}-t_{0})\ket{2}\biggr].
\end{equation}
These two-point correlators can be used to compute four-point correlators.
Out-of-Time ordered correlator (OTOC) \cite{Choudhury:2020yaa, Bhagat:2020pcd} is the 4-point correlating function which can be computed without time-ordering (we can randomly take the points without considering the flow of direction of time). Eq. \eqref{4p} is expressing the `scattering matrix' interpretation of OTOC.
If we use $t'_{1}, t'_{2}, t_{1}$ and $t_{2}$ instead of  $t_{1}, t_{2}, t_{3}$ and $t_{4}$ the OTOC will be:
\be\label{ot1}
Tr\biggr(\hat{q}(t'_{1})\hat{q}(t'_{2})\hat{q}(t_{2})\hat{q}(t_{1})e^{i\hat{H}t_{0}}\ket{0}\bra{0}e^{-i\hat{H}t_{0}}\biggr)=\langle \psi(t'_{1},t'_{2}) | \psi(t_{1},t_{2})\rangle.
\ee
Substituting Eq.\eqref{psizeta}in Eq.\eqref{ot1} one can show that,
\begin{flalign}\label{ot2}
\nonumber \langle \psi(t'_{1},t'_{2}) | \psi(t_{1},t_{2})\rangle &=\frac{\Gamma^{2}(0)}{2}\zeta(t_{1}-t_{0})\zeta^{*}(t'_{1}-t_{0})\biggr[\zeta(t'_{2}-t_{0})\bra{0}+\sqrt{2}\zeta^{*}(t'_{2}-t_{0})\bra{2}\biggr]\biggr[\zeta^{*}(t_{2}-t_{0})\ket{0}+\sqrt{2}\zeta(t_{2}-t_{0})\ket{2}\biggr]\\
&=\frac{\Gamma^{2}(0)}{2}\biggr[\zeta(t_{1}-t_{0})\zeta^{*}(t'_{1}-t_{0})\zeta(t'_{2}-t_{0})\zeta^{*}(t_{2}-t_{0})+\zeta(t_{1}-t_{0})\zeta^{*}(t'_{2}-t_{0})\zeta^{*}(t'_{1}-t_{0})\zeta(t_{2}-t_{0})\nonumber\\
&~~~~+\zeta(t_{1}-t_{0})\zeta(t_{2}-t_{0})\zeta^{*}(t'_{1}-t_{0})\zeta^{*}(t'_{2}-t_{0})\biggr].
\end{flalign}
The lesser Green's function in terms of Heisenberg field operators can be expressed as:
\be \label{lgf}
G_{<}(t_{1},t_{2})=\frac{\sqrt{2}}{\Gamma_{0}^{2}}Tr\biggr(\hat{q}(t_{2})\hat{q}(t_{1})e^{i\hat{H}t_{0}}\hat{\rho} e^{-i\hat{H}t_{0}}\biggr).
\ee
Using the two-point correlator for ground state of the inverted oscillator in Eq.\eqref{2pointa}, we can express the lesser Green's function in the below given form:
\be\label{lgft}
G_{<}(t_{1},t_{2})=\frac{1}{\Gamma_{0}}\zeta(t_{1}-t_{0})\zeta^{*}(t_{2}-t_{0}).
\ee
Using the results of \cite{BenTov:2021jsf} the OTOC in Eq.\eqref{ot2}, can be expressed in terms of lesser Green's function given in Eq.\eqref{lgft} and is given by:
\begin{flalign}\label{o}
\nonumber Tr\biggr(\hat{q}(t'_{1})\hat{q}(t'_{2})\hat{q}(t_{2})\hat{q}(t_{1})e^{i\hat{H}t_{0}}&\ket{0}\bra{0}e^{-i\hat{H}t_{0}}\biggr)\\
&=\frac{\Gamma^{4}_{0}}{2}\biggr[G_{<}(t_{1},t'_{1})G_{<}(t_{2},t'_{2})+G_{<}(t_{1},t'_{2})G_{<}(t_{2},t'_{1})+G_{<}(t_{1},t_{2})G_{<}(t'_{2},t'_{1})\biggr].
\end{flalign}

%========================================

\section{\textcolor{blue}{\textbf{ \large Numerical Results}}}
\label{sec:numerical}
Further explicitly writing the Eq.\eqref{o} by putting the Green's functions from Eq.\eqref{sgf}, one can get the real part of the functional form of out-of-time ordered correlator function, represented as:
\begin{equation}\label{oto}
    \mathcal{F}(t) = \left\{1+2 e^{-f(t)t} cos\left(\frac{m_{0}\Omega(t)q_{0}^{2} t}{2}\right)\right\}\times \frac{1}{8}\left\{m_{0}^{2}\Omega^{2}(t) q_{0}^{4} + f^{2}(t)\right\}.
\end{equation}
From the above equation Eq. \eqref{oto}, it is clear that the exact value of OTOC is sensitive to the functional form of frequency and coupling of oscillator.

In this section, using Eq \eqref{oto} we numerically evaluate the exact results for OTOC measures computed by varying the functional forms of coupling, $f(t)$ and frequency, $\Omega(t)$ of the inverted harmonic oscillator. Particularly we choose the coupling, $f(t)$ and frequency $\Omega(t)$ as quenched parameters and parameterise plots by varying quench protocol. 

We begin by considering the mass, $m(t) = m_{0}e^{\eta t}$, where $m_{0}$ is the initial mass and $\eta$ is positive and real number which measures the rate of increment of mass. The system is very similar to generalized Caldirola-Kanai oscillator. Since, the final form of the Eq \eqref{oto} is independent of $\eta$ we are not very concerned about this parameter. 
We further set the other quantities in Eq. \eqref{oto} as, $m_{0}=1$, $q_{0}=10$. %
%To obtain the constants $A,B$ and $C$ mentioned in Eq. \eqref{expli} we first compute $\sigma_i(t,\delta t)$ and $\gamma_i(t,\delta t)$ at $t\rightarrow0$. Next we, set $d(t\rightarrow0)=f(t\rightarrow0)=0$, defined in, Eq. \eqref{d} and $\sigma_i(0,\delta t)=1$. Using these initial conditions we obtain values of $A,B$ and $C$ which can be inserted in Eq. \eqref{expli}. $\rho_i(t,\delta t)$ and $\gamma_i(t,\delta t)$ are then used to get numerical values of von Neumann and Renyi entanglement entropies, for the aforementioned initial conditions. 
\\

%{\begin{figure}[htb!]
	%\centering	%\includegraphics[width=13cm,he%ight=8cm]{#OTOC4.pdf}
%		\caption{Variation of the OTOC ($\mathcal{F}(t)$) with respect to time $(t)$ for a chosen time-dependent frequency profile and a constant coupling of inverted harmonic oscillator.}
%		\label{fig:samequench}
%\end{figure}

\begin{figure}[!htb]
     	\centering
	    \subfigure[]{\includegraphics[width=8.7cm,height=7.5cm]{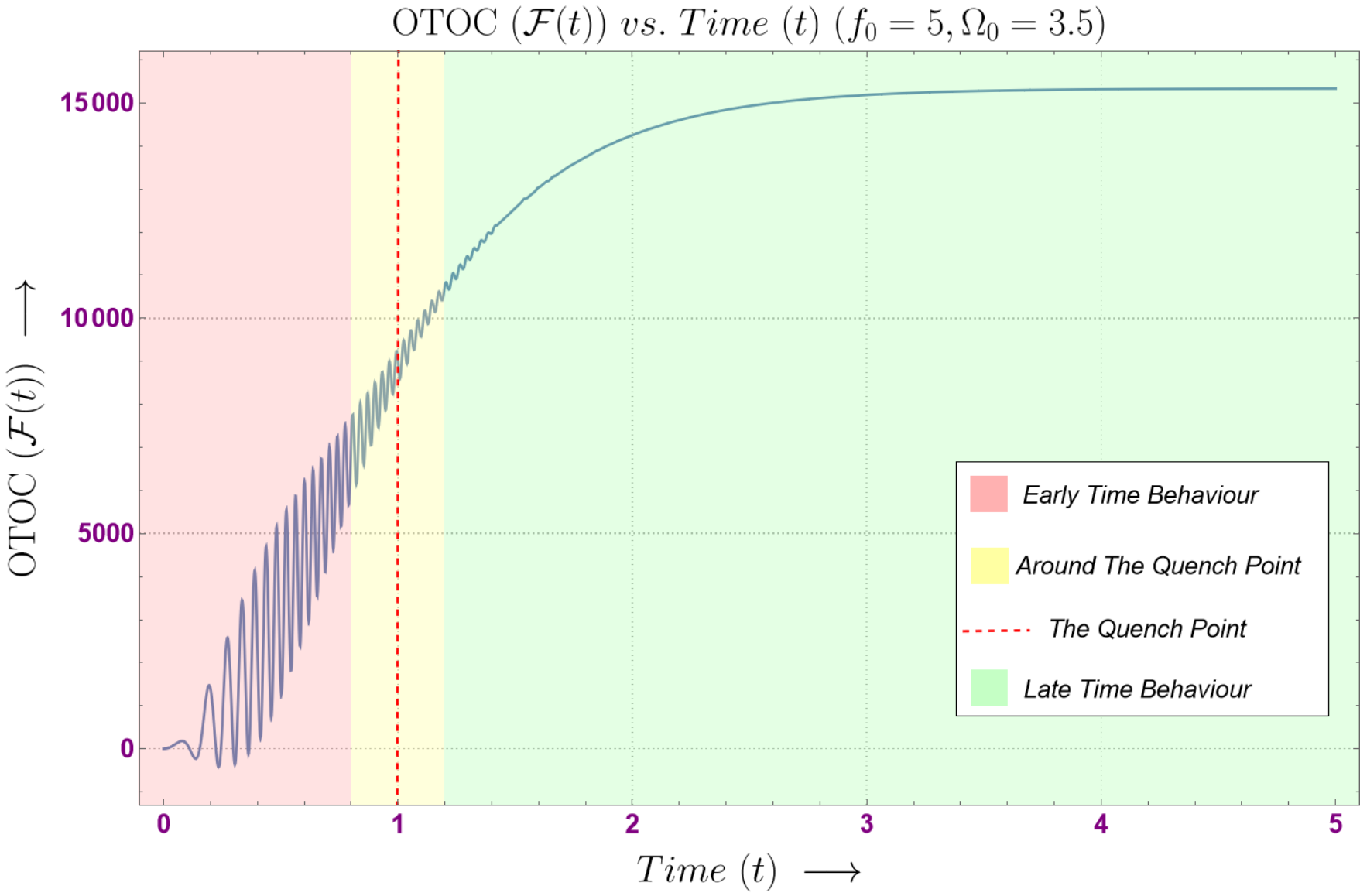}}\label{plota}\hspace{1pt}\hspace{10pt}	\subfigure[]{\includegraphics[width=8.7cm,height=7.5cm]{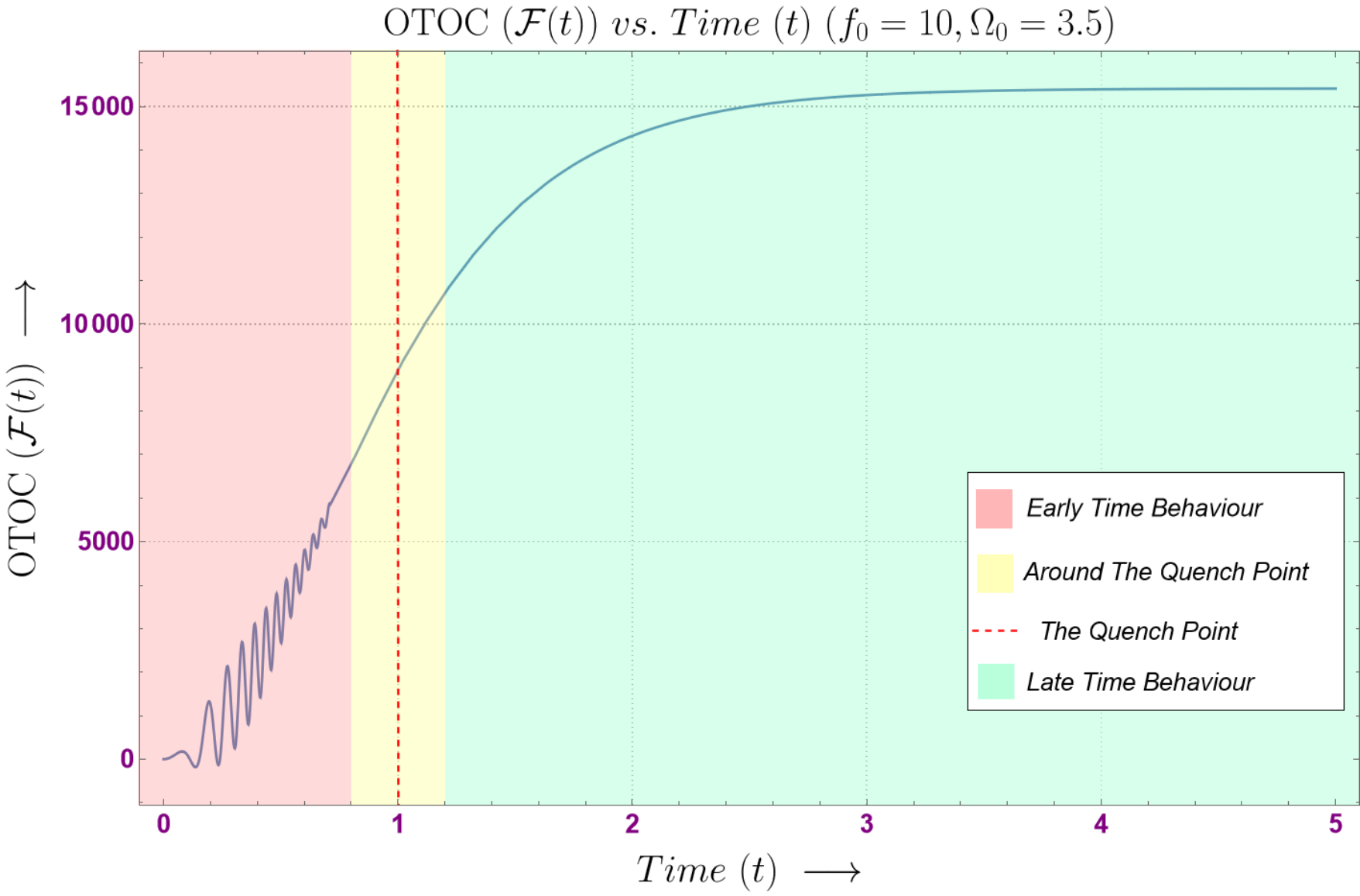}}\label{plotb}\hspace{1pt}%
	    \\\subfigure[]{\includegraphics[width=10.8cm,height=7.6cm]{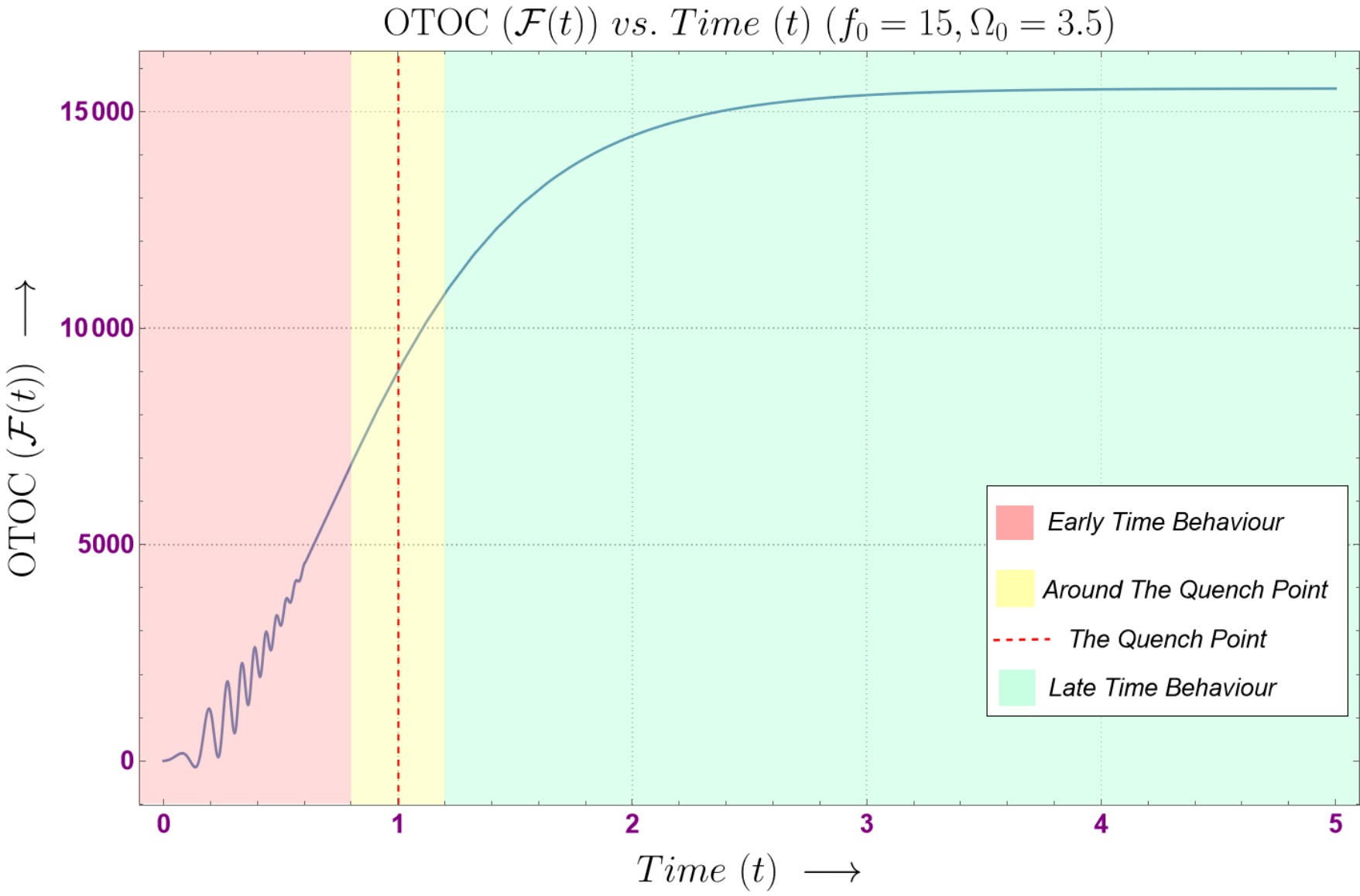}}\label{plotc}\hspace{1pt}
	\caption{Variation of the OTOC ($\mathcal{F}(t)$) versus time $(t)$ for a constant $\Omega_0$ but different $f_0$, such that frequency and coupling of inverted oscillator have same quench profiles.}
	\label{fig:samequench}
\end{figure}

 %%%%%%%%%%%%%%%%%%%%%%%%%%%%%%%%%%%
 %%%%%%%%%%%%%%%%%%%%%%%%%%%%%%%%%%%
 \begin{itemize}
     \item In FIG.\hyperref[fig:samequench]{2}, we show the dynamical behavior of the OTOC, $\mathcal{F}(t)$ for similar quench protocols chosen as the coupling, $f(t)=f_0\tanh(t)$ and frequency, $\Omega(t)=\Omega_0\tanh(t)$ of the inverted oscillator. Here $f_0$ and $\Omega_0$ are free parameters. From all sub plots it is evident that the early time behavior of OTOC is characterised by fluctuating values of OTOC such that the amplitude of these fluctuations decrease as we move further in time. We also observe that for time values on the order of magnitude 1, the OTOC values scale on the order of $10^3$ or $10^4$. Upon approximation, the growth exhibits an exponential behavior of the form:
     $$\mathcal{F}(t) = a e^{(b t)} + c$$ With coefficients which are found by exponential fit of the plot in FIG.\hyperref[plota]{2(c)} to be $a = 1.73 \times 10^4$, $b = 0.442$, and $c = -1.81 \times 10^4$, all of which are non-zero. The apparent linear region near the quench can be attributed to the saturation of the exponential rise that begins after the quench point. At some time beyond the quench point $t=1$, these oscillations are completely damped and we observe a linear rise in the value of OTOC characterising the chaotic nature of the system. Finally, once the dynamics of chosen quench protocols drive system from non-equilibrium to equilibrium, we observe that the OTOC saturates to a constant value irrespective of time. Now, one can choose different triggers to thermalise the dynamical behavior of $\mathcal{F}(t)$ by varying the value of coupling parameter $f_0$. It is evident from the plots that once $f_0$ starts to increase, the quenched coupling triggers, $\mathcal{F}(t)$ in such a way that the fluctuations tend to dampen earlier. 
     \item Particularly in sub plot FIG.\hyperref[plota]{2(a)}, when $f_0=5.0$, we observe rapid oscillations having higher amplitude in value of $\mathcal{F}(t)$ in the region $0<t<0.8$, marked by a red background. As we approach quench point at $t=1$, the amplitude of these oscillations dampens marked by yellow region. The chosen quench protocol then starts to thermalise the system and we observe that the fluctuations completely die out at $t\approx1.6$. The value of $\mathcal{F}(t)$, then rises and saturates for $t>2$. This late time behavior of OTOC is marked by green background.
     \item In sub plot FIG.\hyperref[plotb]{2(b)} we observe a similar dynamical behavior in the value of $\mathcal{F}(t)$. However it is evident that at the chosen value of free parameter $f_0=10$, the fluctuations die out at $t\approx0.8$. Hence the early time behavior of OTOC is characterised by a fluctuating vale of $\mathcal{F}(t)$ marked by red background. This is followed by a rising value of $\mathcal{F}(t)$ for $0.8<t<1.2$, in the region around the quench point marked by a yellow background. Finally beyond the quench point, the rise in the value of $\mathcal{F}(t)$ is followed by thermalisation. This is marked by green background.
     \item In sub plot FIG.\hyperref[plotc]{2(c)}, we increase the free parameter to $f_0=15$. We again observe fluctuations in value of $\mathcal{F}(t)$, although dying away at $t\approx0.5$. Most of the early time behavior of OTOC is characterised by fluctuations which dampen for $0<t<0.8$, marked by a red background. Near to the quench point, in the region $0.8<t<1.2$, we again observe a rising behavior in value of OTOC, marked by a yellow background. The late time behavior evidently shows a trend of thermalisation in the value of $\mathcal{F}(t)$, for the region $t>1.2$, marked by a green background. For this parameters we extract the quantum Lyapunov exponent, by plotting logarithm of $\mathcal{F}(t)$ with respect to time.
     %%lyapunov%%

    \begin{figure}[htb!]
	\centering

	\includegraphics[width=15cm,height=10cm]{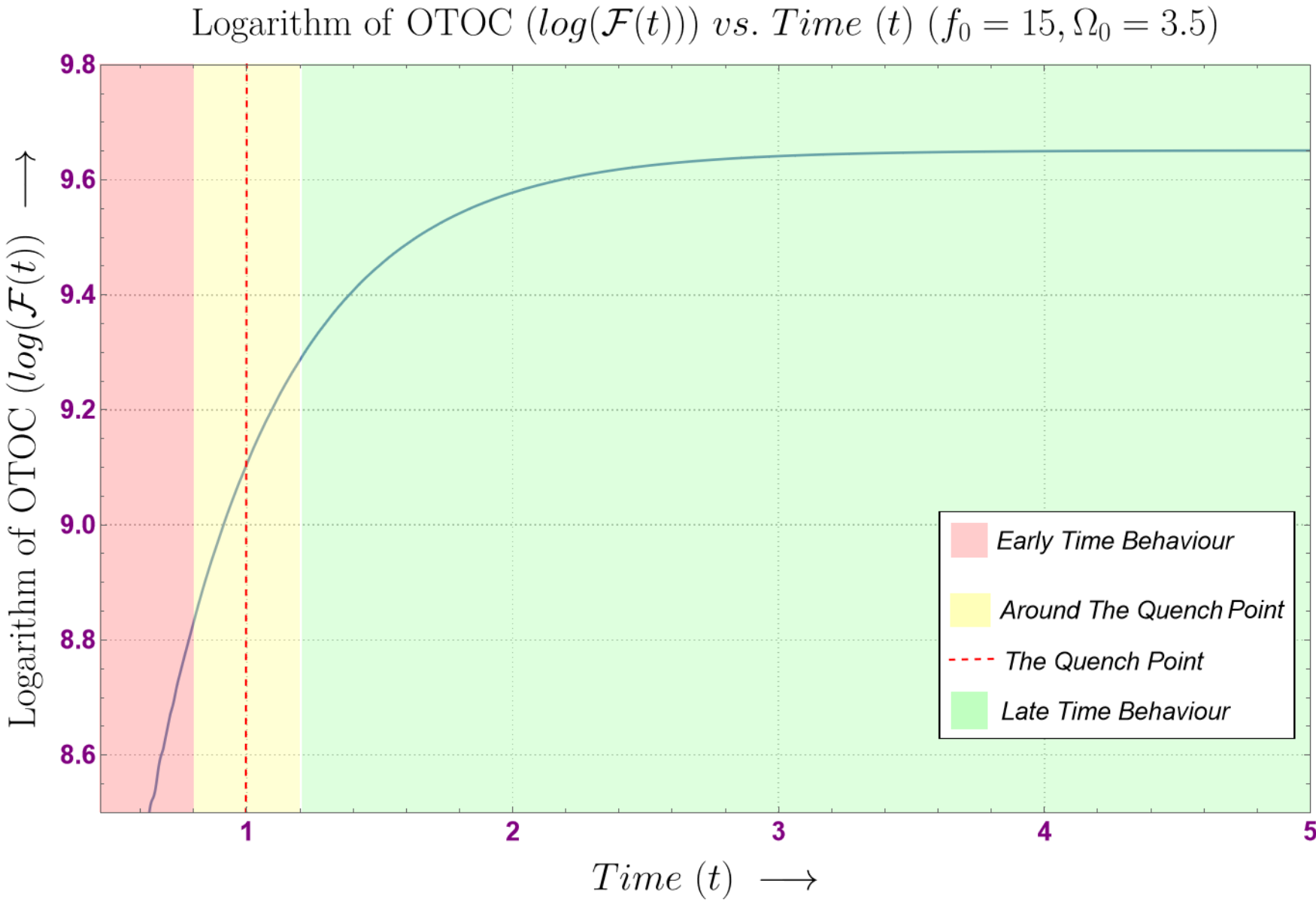}

	\caption{Variation of the Logarithm of OTOC ($\log{\mathcal{F}(t)}$) with respect to time $(t)$ where we choose same quench profiles for frequency and coupling.}
	\label{logOTOC}
    \end{figure}

    \item Quantum Lyapunov exponent, is considered as the rate of growth of the logarithmic value of the OTOC and is often used to describe the chaotic nature of system. Using the value of OTOC from Eq.\eqref{oto}, we can define the Lyapunov exponent for the inverted oscillator as,
    \begin{equation}\label{lyapu}
        \lambda_L=\frac{\partial}{\partial t}\log[\mathcal{F}(t)].
    \end{equation}
    One can focus on the rising trend in $\mathcal{F}(t)$ vs $t$ plot, and approximate the behavior of $\mathcal{F}(t)$ in the rising region as exponential one, $\mathcal{F}(t)\approx e^{\lambda_Lt}$. $\lambda_L$ can then be computed as the slope of the plot of $\log[\mathcal{F}(t)]$ vs $t$.
    \item In FIG.\ref{logOTOC}, we plot the dynamical behavior of the logarithm of OTOC for the same parameters chosen in FIG.\hyperref[plotc]{2(c)}. Then $\lambda_L$ can be computed as  the slope of the rising trend in this figure from $0.8<t<1.2$. The average slope, computed for the nearly linear region of FIG.\ref{logOTOC}, was found to be $\lambda_L=1.124$ while the average rate of change of logarithm of $\mathcal{F}(t)$ (using Eq.\eqref{lyapu}) is $\lambda_L=1.121$. Thus we conclude that the extracted value of Lyapunov exponent using the graph seems to be in agreement with that using the theoretical value. It is evident from the value of $\lambda_L$ that the system of inverted oscillator is chaotic in nature. 
%%%%%%%%%
 \begin{figure}[!htb]
     	\centering
	    \subfigure[]{\includegraphics[width=8.7cm,height=7.5cm]{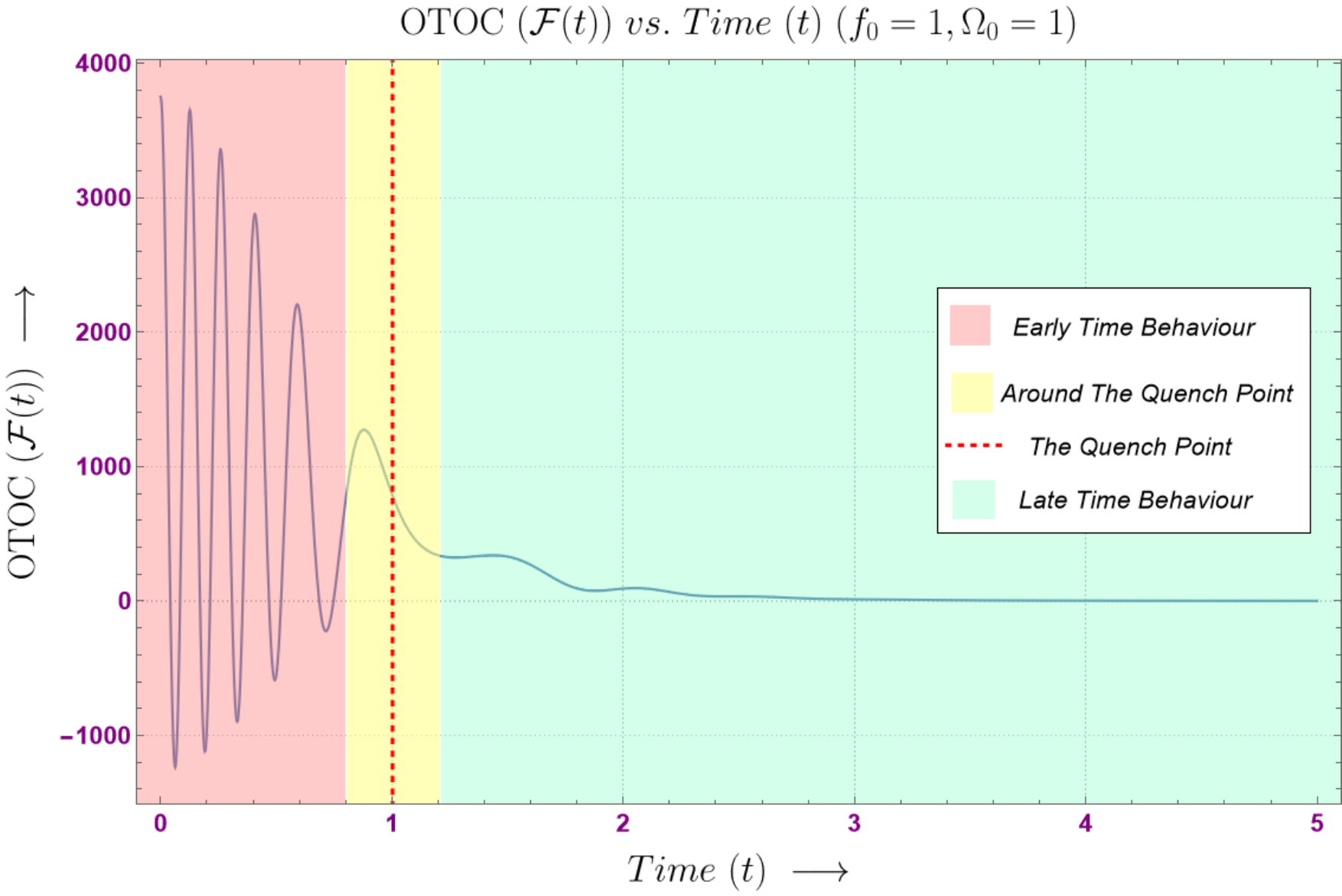}}\label{plot4a}\hspace{1pt}\hspace{10pt}	\subfigure[]{\includegraphics[width=8.7cm,height=7.5cm]{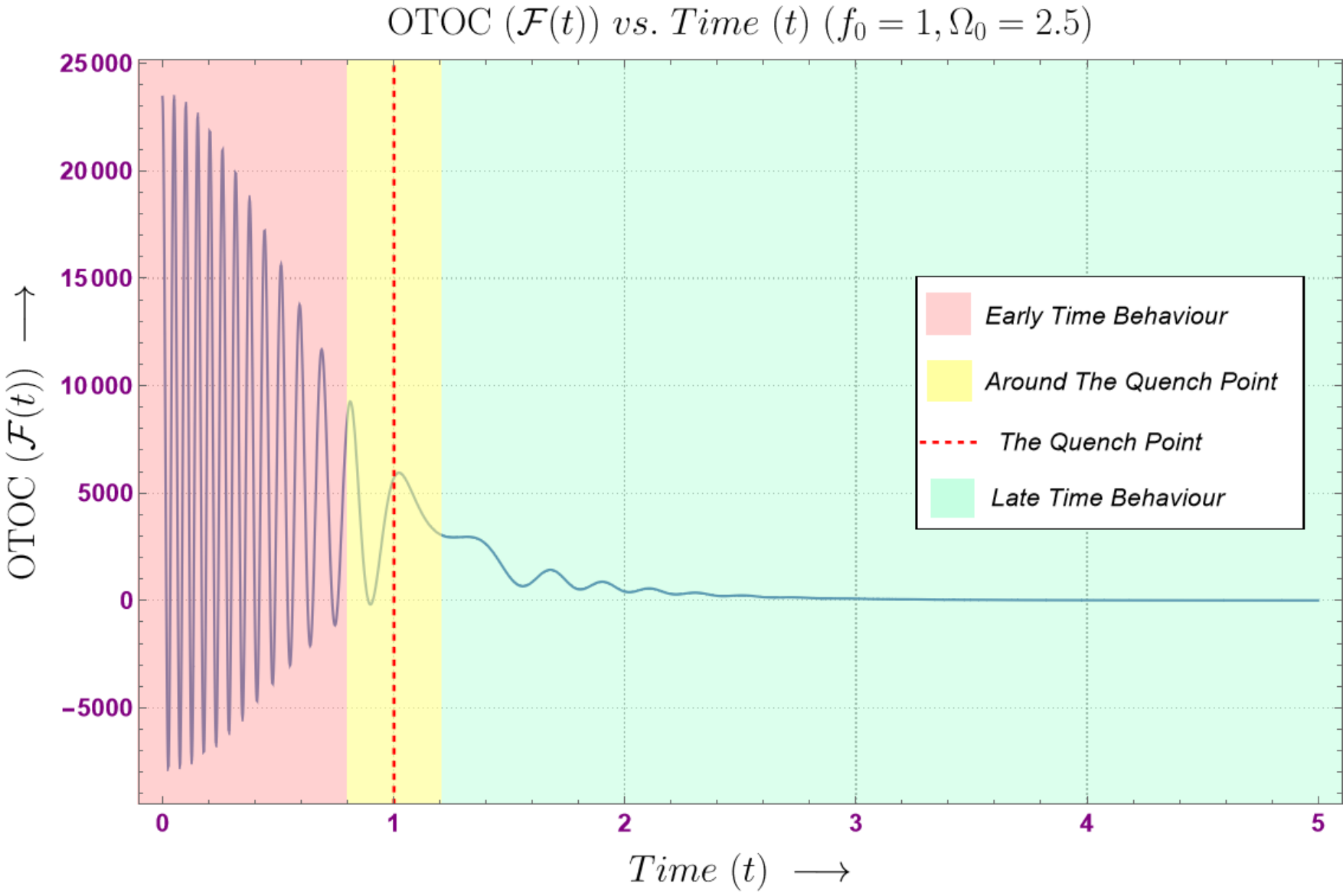}}\label{plot4b}\hspace{1pt}%
	    \\\subfigure[]{\includegraphics[width=10.8cm,height=7.6cm]{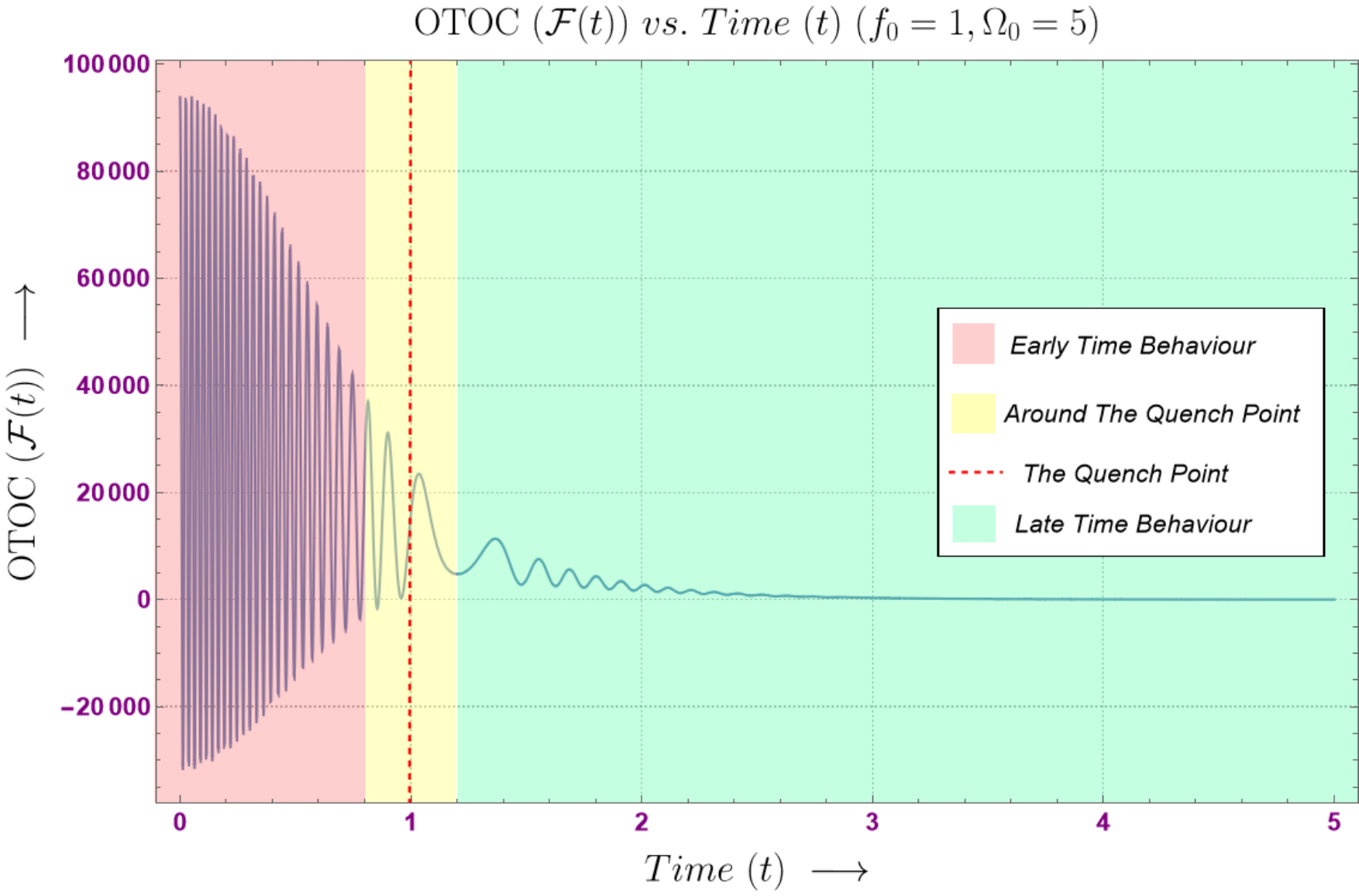}}\label{plot4c}\hspace{1pt}
	\caption{Variation of the OTOC ($\mathcal{F}(t)$) versus time $(t)$ for a constant $f_0$ but different $\Omega_0$, such that frequency and coupling of inverted oscillator have different quench profiles.}
	\label{fig:diffquench}
\end{figure} 
%%%%%%%%%%%%%

     \item In FIG.\ref{fig:diffquench}, we show the dynamical behavior of the OTOC, $\mathcal{F}(t)$ for different quench protocols chosen as the coupling, $f(t)=f_0 ~\text{sech}(t)$ and frequency, $\Omega(t)=\Omega_0\tanh(t)$. For all sub plots it is clear that, at early time the value of $\mathcal{F}(t)$ fluctuates with rapid oscillations which are dampened at some time after the quench point $t=1$. As we move further in time, it is evident that the dynamical effects of quench protocols reduce the amplitude of $\mathcal{F}(t)$ to zero. Also, as we choose increasing values of free parameter, $\Omega_0$, the frequency of fluctuations in the value of OTOC increase.
     \item Particularly in sub plot FIG.\hyperref[plot4a]{4(a)}, when $\Omega_0=1$, we observe that the early time behavior of $\mathcal{F}(t)$ at $0<t<0.8$ is dominated by fluctuations, marked by red background. These oscillations in the value of OTOC tend to dampen as we move further in time. Around the quench point, for $0.8<t<1.2$ we observe a rise, followed by a dip in the value of $\mathcal{F}(t)$, marked by yellow background. At late time, for $t>1.2$, the oscillations tend to die out completely, such that $\mathcal{F}(t)$ becomes zero, this is marked by a green background.
     \item In sub plot FIG.\hyperref[plot4b]{4(b)}, when we increase free parameter as, $\Omega_0=2.5$, we observe rapid oscillations in the value of $\mathcal{F}(t)$ with a higher frequency than that of previous case, in early time region i.e. $0<t<0.8$ (red background). The amplitude of these fluctuations dampens for later times. Around the quench point, $0.8<t<1.2$ and at late times, $t>1.2$  we again observe similar trend to that of previous case, marked by a yellow and green background respectively.
     \item In sub plot FIG.\hyperref[plot4c]{4(c)}, we further increase free parameter, $\Omega_0=5$ and observe a similar trend in the dynamical behavior of $\mathcal{F}(t)$ to that of previous cases. However at early times, for $0<t<0.8$, the fluctuations are more rapid with a higher frequency than that of previous case. This is marked by a red background.
     \item It is evident from the nature of plots, shown in FIG.\ref{fig:samequench}, that the similar quench protocols for coupling and frequency of the inverted oscillator trigger thermalisation such that the OTOC saturates to a non-zero value at very late times.
     On the other hand if we choose different quench protocols, shown in FIG.\ref{fig:diffquench}, for coupling and frequency of the oscillator, the OTOC although having no fluctuation at late times, saturates to zero value.
 \end{itemize}

\section{\textcolor{blue}{\textbf{\large Conclusion}}}\label{discuss}
We have studied the Schwinger-Keldysh path integral formalism in the context of a generic time-dependent inverted oscillator system that is governed by non-equilibrium dynamics, with a generic Hamiltonian and how the formalism can help to generate the out-of-time ordered correlators (OTOCs). The following list of points serves as an epilogue of this paper:
\begin{itemize}
    \item Starting with a time-dependent Hamiltonian of inverted oscillator we use Lewis-Riesenfield invariant method to solve the TDSE and hence compute the eigenstates and eigenvalues. We notice that these eigenvalues are independent of the state $n$, but however are continous funtions of time. Using these eigenstates we have derived the correlation functions in the form of density matrix.
    \item Next we calculated the time-dependent Greens function for the considered Hamiltonian of inverted oscillator. The obtained form of Green's function is used while computing the four point OTOC.
    \item Further, we have derived the expression of time-dependent eigenstates of the Hamiltonian of generalised inverted oscillator in Heisenberg picture. The two-point correlators are then derived by using the form of Green's function. It is evident that the time-dependent OTOC can be expressed in the form of the lesser Green's function.
    \item Next we have expressed the classical action in terms of Heisenberg field operators and Green's functions. This action is then used to compute the generating function of the system. The final form of generating functional is evidently dependent on Green's function and Heisenberg field operators. This generating functional is then used to evaluate the expression for 4-point OTOC.
    \item Since there is no implicit time-ordering in OTOC, one could not use Feynman path integrals to evaluate such correlators. Instead, we have employed the use of Schwinger-Keldysh path integral formalism to compute OTOC for time-dependent inverted oscillator. By choosing a Schwinger-Keldysh contour, we have given the exact form of OTOC for inverted oscillator. Furthermore we also mention the form of influence phase. 
    \item The analytical form of the real part of OTOC is then used to numerically evaluate the dynamical behavior of OTOC. Furthermore, we choose quenched coupling and frequency of the inverted oscillator, so as to comment on the thermalisation of OTOC for different parametric variations.
    \item From the numerical results, it is evident that there are different phases in the dynamical behavior of OTOC, the first being the oscillatory phase where the value of OTOC fluctuates with a decreasing amplitude. The oscillatory phase is dominant at early times.  When we set both the quenched frequency as well as coupling to the same quench protocol, then the quench seems to be thermalising OTOC at late time. However before this thermalisation, clearly we observe a rise in the value of OTOC, showcasing the chaotic nature of the system. We could extract the quantum Lyapunov exponent by using this rise in the value of OTOC. Clearly, the two other phases of OTOC are the exponential rise and saturation. Furthermore we observe that as we increase the coupling free parameter in the chosen quench protocol, the fluctuations stop at earlier time. Hence it can be inferred that increasing coupling essentially make oscillatory phase of OTOC less dominant. Also, the initial conditions when changed have drastic effects on the dynamical behavior of OTOC, again due to the chaotic nature of the system.
    \item On the other hand, if we choose two different quench protocols as the quenched frequency and coupling, then at early time the value of $\mathcal{F}(t)$ fluctuates with rapid oscillations which are dampened at some time after the quench point $t=1$. As there is no rise in the OTOC, one can infer that the system is no longer chaotic. As we move still further in time, it is evident that the dynamical effects of quench protocols though trying to thermalise the value of OTOC, reduce the amplitude of $\mathcal{F}(t)$ to zero.  Also, as we choose increasing values of frequency free parameter, the frequency of oscillations at early time increases rapidly
\end{itemize}
Some future prospects of the present work are appended below point-wise:
\begin{itemize}
\item In this work we have developed to compute the OTOC for  inverted oscillator in presence of quenched parameters using its time dependent ground state.  We can develop the thermal effective field theory of this model and compute the thermal OTOC for the same \cite{BenTov:2021jsf}. Inspired by the work of \cite{Maldacena} we can then use the thermal OTOC to find that our quenched inverted oscillator model exactly saturates to the Maldacena Shenker Stanford (MSS) bound on the quantum {\it Lyapunov exponent} $\lambda\leq 2\pi/\beta$,  where $\beta$ represents the inverse equilibrium temperature of the representative quantum mechanical system after achieving thermalization at the late time scales.  This extended version of the computation will be also extremely helpful to comment on the fact that whether maximal chaos can be achieved from the present set up or not.  Studying the thermal OTOCs and scrambling time scales in systems depending on quenched inverted oscillators can be intriguing.  For that case one need to use the well known {\it thermofield double state} \cite{Cottrell:2018ash} instead of temperature independent ground state to extend the present computation at finite temperature.  One  important assumption we need to use to explicitly perform the computation of OTOC with  {\it thermofield double state} is that the system fully thermalizes at the late time scales.  There are many applications in black hole physics in presence of shock wave geometry \cite{Shenker:2014cwa,Roberts:2014isa,Roberts:2014ifa,Stanford:2014jda,Stanford:2014zyi,Shenker:2013yza,Shenker:2013pqa} and in cosmology \cite{Choudhury:2020yaa,Choudhury:2021tuu} where one can use the derived results of {\it Schwinger-Keldysh formalism} as well as OTOC.  Though the specific effect of the time dependent quench in the coupling parameters of the underlying theoretical set up have not been studied yet in both the above mentioned problems.  It would be really interesting to explicitly study these effects within the framework of black hole physics as well as in cosmology.

\item  Since in this computation the time dependent quantum quench is playing significant role,  it is very natural to study the thermalization phenomenon explicitly from the present set up,  as we all know quantum quench triggers the thermalization procedure in general.  If the system under consideration fully thermalizes at the late time scales then the corresponding system can be described in terms of pure thermal ensemble and the quantum states are described by thermofield double states.  But this type of features are very special and can only happen in very specific small classes of specific theories.  If we have a slow/fast quench in the coupling parameters of the theory then it is naturally expected from set up that before quench,  just after quench and at late time scales the corresponding quantum mechanical states will be different.  In technical language these are pre quench,  post quench-Calabrese Cardy (CC) or generalized Calabrese Cardy (gCC) \cite{Banerjee:2019ilw,Mandal:2015kxi,Paranjape:2016iqs,Banerjee:2021lqu} and Generalized Gibbs ensemble (if not fully thermalizes) \cite{Banerjee:2019ilw,Mandal:2015kxi,Paranjape:2016iqs,Banerjee:2021lqu} or pure {\it thermofield double} states.  This particular information is extremely important because the computation of OTOC is fully dependent on the underlying quantum states.  It is expected that for a given theoretical set up we need to use three different quantum states in three different regimes to compute the expression for OTOC in presence of time-dependent quench in the coupling parameters of the theory.  In the OTOC computation, which we have performed in this paper,  we have used the ground states at zero temperature,  which is technically a pre quench state.  It would be really great if one could study the OTOC using the post quench CC/gCC as well as Generalized Gibbs ensemble/thermal states.  Now one very crucial point is that,   which people usually don't use to construct these states at different regimes,  which is the {\it invariant operator method}.  This method allows us to construct all possible states at different time scales having a quench profile in the respective parameters of the theory.  This implies that the {\it invariant operator method}.  automatically allows us to construct the corresponding quantum states in the pre quench and post quench region,  which is necessarily needed to compute OTOC.  Most importantly,  {\it invariant operator method} allows us to construct the quantum mechanical states at each instant of time in the time scale,  except at very late time scale when the full thermalization or the effective thermalization has been achieved by the underlying physical set up.

\item The OTOC has been used to understand the dynamics of cosmological scalar fields \cite{Choudhury:2020yaa}, it would be interesting to explore such ideas in the presence of quench. Our work deals with inverted oscillators in presence of a quench which can be applied to many models appearing in cosmology, quantum gravity and condensed matter physics. 
\end{itemize}

%\end{widetext}

\textbf{Acknowledgement:}
~~ The research work of SC is supported by  CCSP, SGT University, Gurugram Delhi-NCR by providing the Assistant Professor (Senior Grade) position.  SC also would like to thank CCSP, SGT University, Gurugram Delhi-NCR for providing the outstanding work friendly environment.  We would like to thank Silpadas N from Department of Physics,  Pondicherry University for the collaboration in the initial stage of the work.  SC would like to thank all the members of our newly formed virtual international non-profit consortium Quantum Aspects of the Space-Time \& Matter (QASTM) for for elaborative discussions. Last but not least, we would like to acknowledge our debt to the people belonging to the various part of the
world for their generous and steady support for research
in natural sciences.

\onecolumngrid
%\newpage
\appendix
\section{\textcolor{blue}{\textbf{\normalsize Integration over initial field configuration}}}\label{intif}
~~~~In this appendix our prime objective is to simplify the expression for wavefunction in Eq.\eqref{psif} by evaluating the contribution of integral over initial field configuration, $q_0$. We begin with writing the contribution of initial field to the wavefunction of inverted oscillator, $\chi(q_f|J)$ as:\\
\begin{equation}
\begin{aligned}\label{chi}
\chi(q_f|J)=\int_{-\infty}^{\infty}~dq_{0}\exp\Biggr[\frac{i}{2G(T)}\Bigg(\Biggl[\biggr\{f(t_0)t_0)+\frac{i}{x_0^2}\biggr\}G(T)+m(t_0)\dot{G}(T)+\frac{\biggr[\frac{m^2(t_1)+m^2(t_0)}{m(t_1)}\biggr]G(t_f-t_1)}{2G(t_1-t_0)}\Biggr]q_0^2\\-q_0\biggr[[m(t_f)+m(t_0)]q_f-[m(t_0)+1]\int_{t_0}^{t_f}dtG(t_f-t)J(t)+\biggr[m(t_1)-\frac{m(t_0)m(t_f)}{m(t_1)}\biggr]q_f\biggr]\Bigg)\Biggr].
\end{aligned}
\end{equation}
Further we define,
\be \label{at}
\mathscr{A}(t)=\biggr[f(t_0)m(t_0)+\frac{i}{x_0^2}\biggr]G(T)+m(t_0)\dot{G}(T)+\frac{\biggr[\frac{m^2(t_1)+m^2(t_0)}{m(t_1)}\biggr]G(t_f-t_1)}{2G(t_1-t_0)}.
\ee
Here, 
\be
\frac{1}{x_{0}^{2}}=\frac{im(t)}{2K(t)}(\dot{K}(t)-f(t)K(t)).
\ee
Using Eq.\eqref{at} we can rewrite the expression for Eq. \eqref{chi} as:
\be
\begin{aligned}\label{chiJ2}
\chi(q_f|J)=&\exp{\biggr[\frac{-i\mathscr{A}^{-1}(t)}{8G(T)}\biggr(\{m(t_f)+m(t_0)\}q_f -[m(t_0)+1]\int_{t_0}^{t_f}dt~G(t_f-t)J(t)+\biggr[m(t_1)-\frac{m(t_0)m(t_f)}{m(t_1)}\biggr]q_f\biggr)^2\biggr]} \times \\&\int_{-\infty}^{\infty}dq_0\exp{\left(\frac{i\mathscr{A}(t)}{2G(T)}[q_0-\beta]^2\right)},
\end{aligned}
\ee
where the factor $\beta$ is,
\be
\begin{aligned}
\beta=\biggr\{\frac{[m(t_f)+m(t_0)]q_f-[m(t_0)+1]\int{t_0}^{t_f}dt~G(t_f-t)J(t)+\biggr[m(t_1)-\frac{m(t_0)m(t_f)}{m(t_1)}\biggr]q_f}{2A(t)}\biggr\}.
\end{aligned}
\ee
Then performing the integration over $q_0$ one can rewrite the Eq. \eqref{chiJ2} as:
\be\label{xif}
\begin{aligned}
\chi(q_f|J)=\sqrt{2\pi i G(T)\mathscr{A}^{-1}(t)}\exp\biggr[\frac{-i\mathscr{A}^{-1}(t)}{8G(T)}\biggr(\biggr\{m(t_f)+m(t_0)+\biggr[m(t_1)-&\frac{m(t_0)m(t_f)}{m(t_1)}\biggr]\biggr\}q_f\\&-\big(m(t_0)+1\big)\int_{t_0}^{t_f}dt~G(t_f-t)J(t)\biggr)^2\biggr].
\end{aligned}
\ee
Since we have integrated over the initial field configuration, the generating function for inverted oscillator gets simplified and only depends on the final field configuration, $q_0$ as shown in the section \ref{genfuncsk}.

\section{\textcolor{blue}{\textbf{ \large Calculation of Green's Function}}}\label{greenf}
In this work we have obtained the generating function and OTOC for inverted oscillator, in terms of Green's function. We compute this Green's function for inverted oscillator in this appendix.
\noindent Although there are several time-dependent Green's functions, for our system the most important is the outgoing or retarded Green's function. Let $U(t,t')$ is the time operator which depends on final time $t'$ and initial time $t$. Since the Hamiltonian is time-dependent for our system, $U(t,t')$ will satisfy the TDSE. We can write it as,
\be
i\frac{\der{U(t,t')}}{\der{t}}=\hat{H}(t)U(t,t').
\ee
When we put the Hamiltonian of inverted oscillator, Eq.\eqref{IO} in the above equation we get,
\be
\begin{aligned}
\displaystyle
i \frac{\der{U(t,t')}}{\der{t}}=-\frac{1}{2m(t)}\frac{\der^2{U(t,t')}}{\der{q^2}}\displaystyle-\frac{1}{2}m(t)\Omega^2(t)q^2(t)U-\frac{i}{2}f(t)U-\frac{i}{2}f(t)q(t)\frac{\der{U}}{\der{q}}.\\
\end{aligned}
\ee
From the above equation,
\be
i\frac{\der{U}}{\der{t}}=-\frac{1}{2}m(t)\Omega^2(t)q^2(t)U-\frac{i}{2}f(t)U.
\ee
After multiplying the above equation by $i$ and re-arranging, we obtain the following differential equation:
\be
\frac{\der{U}}{\der{t}}+\biggr\{\frac{1}{2}f(t)-\frac{i}{2}m(t)\Omega^2(t)q^2(t)\biggr\}U=0.
\ee
The integrating factor for this differential equation will be $e^{\int\{\frac{1}{2}f(t)-\frac{i}{2}m(t)\Omega(t)q^2(t)\}dt}$. The solution of the above differential equation can be expressed using integrating factor as,
\begin{equation}\label{UG}
U(t)=c~e^{\frac{1}{2}\int\left\{im(t)\Omega^{2}(t)q^{2}(t)-f(t)\right\}dt}.
\end{equation}
Here $c$ is a real constant.
In the quantum mechanics, the outgoing or retarded Green's Function is defined by,
\be
G_+(q,t;q',t')=\Theta(t-t')\bra{q}U(t,t')\ket{q'},
\ee
and the incoming or advanced Green's function is:
\be
G_-(q,t;q',t')=\Theta(t'-t)\bra{q}U(t,t')\ket{q'}.
\ee
The combined form of Green's function using Eq.\eqref{UG} can be given as: 
\bea\label{gfint}
\hat{G}_\pm&=&\pm \Theta\biggl(\pm(t-t')\biggr)U(t,t')\nonumber\\&=&\pm\Theta\biggr(\pm(t-t')\biggr)c~e^{\frac{1}{2}\int\left\{im(t)\Omega^2(t)q^2(t)-f(t)\right\}dt}.
\eea
In above equation, we can eliminate the integration from the power of the exponential by assuming,
\begin{equation}\label{Gam}
    \frac{d(\Gamma(t)t)}{dt}=\frac{1}{2}(im(t)\Omega^2(t)q^2(t)-f(t)).
\end{equation}
Using Eq.\eqref{Gam}, one can rewrite the Green's function in Eq.\eqref{gfint} as:
\bea\label{sgf}
G(t)&=& c~e^{\int\frac{d(\Gamma(t)t)}{dt}~dt}=~c~e^{\Gamma(t)t}.
\eea
For applying the boundary condition we need to change the Eq.\eqref{sgf}~in terms of hyperbolic sines and cosines as given below,
\be
G(t)= A~\sinh{(\Gamma(t)t)}+B~\cosh{(\Gamma(t)t)}.
\ee
Applying the boundary conditions $G(0)=0$ and  $\dot{G}(0)=1$, the Green's function for inverted oscillator is,
\be\label{gfhyp}
G(t)=\frac{1}{\Gamma_{0}}~\sinh{\biggr(\Gamma(t)t\biggr)}.
\ee

\bibliography{referencesnew}
\bibliographystyle{utphys}

\end{document}